%% file: main.tex
\documentclass{article}
\usepackage{amsmath,amssymb,amsfonts}
\usepackage{bbold}
\usepackage{booktabs,multirow}
\usepackage{array}
\usepackage{makecell, cellspace}
\usepackage{diagbox}
\usepackage[dvipsnames, table]{xcolor}
\usepackage{colortbl}
\usepackage{youngtab}
\usepackage{ytableau}
\usepackage{tikz}
\usepackage{siunitx}
\usepackage[a4paper, total={6.5in, 8.5in}]{geometry}
\usepackage{comment}
\usepackage[compat=1.1.0]{tikz-feynman}
\usepackage[hidelinks]{hyperref}
\usepackage[bottom]{footmisc}
\usepackage{slashed}
\usepackage{authblk}
\usetikzlibrary{calc}
\usetikzlibrary{fadings,decorations.pathmorphing}

\newcolumntype{C}{>{$}c<{$}}

\makeatletter
\newif\iftikz@shading@path
\tikzset{
    fading sep/.store in=\fadingsep,
    fading sep=0.25cm,
    shaded path/.code={%
        \iftikz@shading@path%
        \else%
            \tikz@shading@pathtrue%
            \tikz@addmode{%
                \pgfinterruptpicture%
                    \begin{tikzfadingfrompicture}[name=.]
                        \path [shade=none,fill=none]#1;%
                        \xdef\fadingboundingbox{{\noexpand\pgfpoint{\the\pgf@picminx-\fadingsep}{\the\pgf@picminy-\fadingsep}}%
                            {\noexpand\pgfpoint{\the\pgf@picmaxx+\fadingsep}{\the\pgf@picmaxy+\fadingsep}}}%
                        \expandafter\pgfpathrectanglecorners\fadingboundingbox%
                        \pgfusepath{discard}%
                    \end{tikzfadingfrompicture}%
                \endpgfinterruptpicture%
                \expandafter\pgfpathrectanglecorners\fadingboundingbox%
                \def\tikz@path@fading{.}%
                \tikz@mode@fade@pathtrue%
                \tikz@fade@adjustfalse%
                \pgfpointscale{0.5}{\expandafter\pgfpointadd\fadingboundingbox}%
                \def\tikz@fade@transform{shift={(\the\pgf@x,\the\pgf@y)}}%
            }%
        \fi%
    }
}

\newcommand{\NDC}{{N_{\text{DC}}}}
\newcommand{\LDC}{{\Lambda_{\text{DC}}}}
\newcommand{\MGUT}{{M_{\text{GUT}}}}
\newcommand{\SM}{{\text{SM}}}

\newcommand{\invalpha}[2]{\alpha_{#1}^{-1}\left(#2\right)}

\newcommand{\one}{\mathbf{1}}
\newcommand{\two}{\mathbf{2}}
\newcommand{\three}{\mathbf{3}}
\newcommand{\four}{\mathbf{4}}

\newcommand{\seven}{\mathbf{7}}

\newcommand{\threebar}{\mathbf{\bar{3}}}
\newcommand{\fivebar}{\mathbf{\bar{5}}}
\newcommand{\five}{\mathbf{5}}
\newcommand{\ten}{\mathbf{10}}
\newcommand{\fifteen}{\mathbf{15}}
\newcommand{\twentyfour}{\mathbf{24}}

\newcommand{\lag}{{\mathcal{L}}}

\newcommand\scalemath[2]{\scalebox{#1}{\mbox{\ensuremath{\displaystyle #2}}}}

\newcommand{\be}{\begin{equation}}
\newcommand{\ee}{\end{equation}}
\usepackage[
backend=biber,
sorting=none
]{biblatex}
\renewcommand{\tilde}{\widetilde}

\title{Models of Accidental Dark Matter with a Fundamental Scalar}

\author[a]{Stefano Palmisano}
\author[b]{Francesco Rescigno}
\author[c]{Federica Troni}
\affil[a]{Dipartimento di Fisica, Sapienza Università di Roma and INFN, Sezione di Roma, Piazzale Aldo Moro 2, Roma, Italy}
\affil[b]{Dipartimento di Fisica, Università degli studi di Roma "Tor Vergata" and INFN, Via della Ricerca Scientifica 1, Roma, Italy}
\affil[c]{Dipartimento di Fisica, Sapienza Università di Roma, Piazzale Aldo Moro 2, Roma, Italy}
\makeatletter
\def\input@path{{composite-dm/}{model-building/}{pheno/}{gut/}}
\makeatother

\DeclareSIUnit\barn{b}
 
\begin{document}

\fontsize{12pt}{14pt}\selectfont

\maketitle

\begin{abstract}
     We consider models of accidental dark matter, namely models in which the dark matter is a composite state that is stable thanks to an accidental symmetry of the theory. The fundamental constituents are vectorlike fermions, taken to be fragments of representations of the grand unifying gauge group $SU(5)$, as well as a scalar singlet. All the new fields are charged under a new confining gauge group, which we take to be $SU(N)$, leading to models with complex dark matter. We analyse the models in the context of $SU(5)$ grand unification with a non-standard approach recently proposed in the literature. The advantage of including the scalar mainly resides in the fact that it allows 
     several undesired accidental symmetries
     to be broken, 
     leading to a larger set of viable models with respect to previous literature, in which only fermions (or only scalars) were considered. Moreover these models present distinct novelties, namely dark states with non-zero baryon and lepton number and the existence of composite \emph{hybrid} states of fermions and scalars. We identify phenomena that are specific to the inclusion of the scalar and discuss possibilities to test this setup.
\end{abstract}

\pagebreak

\tableofcontents

\pagebreak
 
\section{Introduction}
\input{intro}

\section{General Aspects of the Models} \label{sec:outline}
\input{CDM_outline}

\section{Model Classification} \label{sec:classification}
\input{classification}

\section{\texorpdfstring{$SU(5)$}{SU(5)} Grand Unification}\label{sec:GUT}
\input{gut}

\section{Phenomenology} \label{sec:pheno}
\input{pheno}

\section{Conclusions and Outlook} \label{sec:conclusions}
\input{conclusions}

\section*{Acknowledgements}

Special thanks go to Roberto Contino for suggesting this work and guiding us thoroughly throughout its course. We are grateful to  Marco Nardecchia for insightful discussions on the phenomenological aspects of our models. SP would like to thank Filippo Nardi for useful hints on some of the calculations.

\pagebreak
  
\appendix
\section{Examples of Model Calculations}\label{app:examples}
\input{model-building/examples}
\section{Extending Models}
\input{model-building/Minimal_extension}\label{app:extension}
\section{Full List of Viable Models}\label{app:models}

\input{model-building/all-models}
\clearpage
\printbibliography
\end{document}

%% file: intro.tex
The exact nature of dark matter (DM) is yet unknown, and a great effort has been made on the theoretical side to imagine and explore a variety of possible scenarios, ranging from extended objects such as primordial black holes \cite{PBH} to modifications of gravity \cite{MOND}. However, the most extensively studied option to solve this puzzle is that in which the DM is comprised by new particles, whose relic abundance is set by various mechanisms
. Much work has been devoted to special cases in which the DM puzzle may be solved in combination with other fundamental issues of the the standard model of particle physics (SM). Two examples are axions, which are perfectly suitable DM candidates \cite{axiondm}, or supersymmetry, which naturally provides for a DM candidate \cite{SUSYDM}. In other cases one can introduce new particles 
or entire new sectors 
specifically to address the DM puzzle, perhaps gaining insights or solutions for other issues on the side.

One key feature that any theory of DM must present is an explanation of why the DM is stable on cosmological time scales, as required by observations. In our universe, there are other forms of matter which are stable on such long time scales, namely ordinary protons and electrons. Indeed one of the greatest successes of the SM is the understanding of the observed global symmetries in terms of \emph{accidental} symmetries of the Lagrangian at renormalizable level. These symmetries are the baryon number and the three lepton flavors, and they guarantee the stability of the lightest charged states. 

These symmetries are called accidental since they are not symmetries of the theory if one considers higher dimensional operators in the Lagrangian, which are suppressed by increasing powers of the energy scale of ultraviolet (UV) physics. In other words, they are only symmetries of the theory at low enough energies when the contributions from these operators can be ignored, analogously to the spherical symmetry of the electric field produced by a charge distribution whose size is much smaller than the distance from which it is observed, so that higher multipoles are irrelevant.

In this work we consider extensions of the SM that can solve the DM puzzle with new particles that are stable thanks to an accidental symmetry of the theory. In order to do so, we introduce a non-Abelian dark sector that undergoes confinement. 
Its accidental symmetries determine which of the low energy states (dark hadrons) are stable and can play the r\^{o}le of DM.
This way of tackling the DM puzzle has been given serious consideration in the literature (among many others \cite{stealthdm, Kribs:review-composite, Harigaya, dmnaturalness,Contino:2020, CDM,WCD, Redi:dark-nuclei, Mahbubani:dark-nucleosynthesis, gluequark}). The advantage is that such theories are UV complete, and that the desired properties of the DM descend directly and solely from the quantum numbers of the fundamental constituents and the ensuing accidental symmetries. Moreover, the richness of their low energy spectrum makes them especially promising from a phenomenological standpoint, especially at colliders, since generally these dark partners carry SM charges and may be light enough to be produced, and in cosmology, where they may realize various non-standard scenarios.

In classifying the models, we follow \cite{CDM}, in which the SM was extended with fermionic \emph{dark quarks} (Dq) in the fundamental representation of a new \emph{dark color} (DC) gauge group (they considered both $SU(\NDC)$ and $SO(\NDC)$ groups; we consider only the former). These new fields transformed in vector-like representations of the SM gauge group, in such a way that, unlike in technicolor models or even in quantum chromodynamics (QCD), the confinement does not break the electroweak gauge symmetry. This scenario is known in the literature as vector-like confinement \cite{VLC, VLC2}. In \cite{CDM} only a relatively small number of models were found to be viable. This was due essentially to difficulties in breaking the large number of undesired accidental symmetries that the models generally possess. Indeed any stable symmetry leads to the stability of a dark state, which is in general electrically charged and/or colored, rendering models in which they arise unacceptable. We are interested in grand unification as a criterion for the selection of the models. With this further restriction, essentially one model was found to be acceptable in \cite{CDM}. The novelty of our work in this sense is twofold. On the one hand, along the fermionic Dqs, we consider a scalar field transforming in the fundamental representation of $SU(\NDC)$ but as a singlet under the SM gauge group\footnote{More complicated representations for this scalar may of course be considered.}. On the other, following \cite{verma}, we relax the criterion for grand unification, which results in a more generous selection of models.

The presence of light elementary scalars notoriously introduces what is known as hierarchy problem, as is also very famously the case for the SM.
We postpone entirely the discussion on the hierarchy problem that arises here, as we consider a dark fundamental scalar, assuming that a mechanism exists that stabilizes its mass -- and possibly that of the Higgs boson as well. 
We focus instead on several advantages provided by this setup. 
First of all we show how, by allowing new kinds of interactions between the SM and the dark sector, a larger set of viable models is found. Furthermore, Dqs acquire baryon or lepton numbers depending on their SM representations, and new types of bound state arise, made either of only scalars\footnote{Such a possibility was already explored in \cite{strumia-scalardm}.  In that work, no fundamental fermions were considered, and the interesting \emph{complementarity} between the higgsed and confined phases of non-Abelian theories exploited \cite{Fradkin-Shenker, tHooft-duality, dim-raby-suss}. We leave the treatment of this duality in the context of the models we here propose to future works.} or both of scalars and fermions. In light of these peculiarities we discuss the phenomenological consequences of the presence of the scalar in accidental composite dark matter models.

The rest of this work is organized as follows. In Sec. \ref{sec:outline} we discuss the generalities of accidental composite DM, reviewing the content of \cite{CDM} and highlighting the novelties of our setup, with some examples. We discuss various possible mass orderings and the scenarios they produce. In Sec. \ref{sec:classification} we furnish a full classification of the models. In Sec. \ref{sec:GUT} we analyse the models in the context of $SU(5)$ grand unification, employing the relaxed criterion proposed in \cite{verma}. In Sec. \ref{sec:pheno} we discuss aspects of phenomenology, with focus on the impact of the dark scalar. In Sec. \ref{sec:conclusions} we summarize and discuss our results.

%% file: composite-dm/CDM_outline.tex
\begin{table}[!b]
\caption{Left-handed Dqs are taken to be fragments of $SU(5)$ representations (first column), decomposed under the SM gauge groups (middle columns). The rightmost columns show the contributions of each of the Dqs to the three SM $\beta$-functions. Each of the reported values is to be multiplied by $2 \NDC$, the Dqs being fundamentals of $SU(N)_{\text{DC}}$ and vectorlike with respect to the SM. Last row: SM singlet scalar Dq. }
    \label{tab:reps}
    \vspace{2mm}
    \renewcommand{\arraystretch}{1.3}
    \centering
    $\begin{array}{|c|c|c|c|c|c|c|c|}
    \hline
    SU(5) & \text{Name} & SU(3)_c & SU(2)_L & U(1)_Y & \Delta b_3 & \Delta b_2 & \Delta b_Y \\
    \hline
    \rowcolor{yellow!25}\one   & N & \one & \one & 0 & 0 &0 &0 \\
    \hline
    \rowcolor{cyan!25}& D & \threebar & \one  & \frac{1}{3} & \frac{1}{3} & 0 &    \frac{2}{9} \\
    \rowcolor{cyan!25}\multirow{-2}{*}{$\fivebar$} & L & \one & \two & -\frac{1}{2} & 0 & \frac{1}{3} & \frac{1}{3} \\
    \hline
    \rowcolor{orange!25} & U & \threebar & \one & -\frac{2}{3} & \frac{1}{3} & 0 & \frac{8}{9} \\
    \rowcolor{orange!25}& E & \one & \one & 1 & 0 & 0 & \frac{2}{3} \\ 
    \rowcolor{orange!25}\multirow{-3}{*}{$\ten$}& Q & \three & \two & \frac{1}{6} & \frac{2}{3} & 1 & \frac{1}{9} \\
    \hline
    \rowcolor{green!25}\multirow{3}{*}{$\fifteen$} & Q & \three & \two & \frac{1}{6} & \frac{2}{3} & 1 & \frac{1}{9} \\
    \rowcolor{green!25}& T & \one & \three & 1 & 0 & \frac{4}{3} & 2 \\ 
    \rowcolor{green!25}\multirow{-3}{*}{$\fifteen$}& S & \mathbf{6} & \one & -\frac{2}{3} & \frac{5}{3} & 0 & \frac{8}{9} \\
    \hline
    \rowcolor{purple!25} & V & \one & \three & 0 & 0 & \frac{4}{3} & 0 \\
    \rowcolor{purple!25} & G & \mathbf{8} & \one & 0 & 2 & 0 & 0 \\
    \rowcolor{purple!25}& X & \threebar & \two & \frac{5}{6} &  \frac{2}{3} & 1 & \frac{25}{9} \\
    \rowcolor{purple!25}\multirow{-4}{*}{$\twentyfour$}& N & \one & \one & 0 & 0 &0 &0 \\
    \hline
    \rowcolor{yellow!25}\one & \phi & \one & \one & 0 & 0 & 0 & 0 \\
    \hline
    
    \end{array}$   
\end{table}
We extend the SM with a dark sector containing several Dqs in the fundamental representation of a new $SU(\NDC)$ gauge symmetry. In all the models we consider a scalar Dq $\phi$ which is a total singlet under the SM gauge group\footnote{We use the notation $\left(\mathcal{R}_c, \mathcal{R}_L\right)_Y$ for a field transforming under the SM gauge group $SU(3)_c \otimes SU(2)_L \otimes U(1)_Y$\,.}
\be
\phi \sim  (\one,\one)_0\,.
\ee
We also consider a number of fermionic Dqs. Following \cite{CDM}, since we wish to study these extensions in the context of a $SU(5)$ grand unification scheme, we consider fermionic Dqs belonging to fragments of the lowest $SU(5)$ representations, as shown in Table \ref{tab:reps}. We assume that they are much lighter than their GUT partners, which we assume to not come into play in the cosmological history because they were never populated, being heavier than the reheating temperature. We name the models by just their light fermionic Dq content, always implicitly considering the dark scalar to be light. Since the models are taken to be vector-like with respect to the SM, if we say that a model contains the left-handed Dq field $\Psi$, we implicitly mean that another field $\Psi^c$ is light, which is a left-handed field in the anti-fundamental of $SU(\NDC)$, belonging to the conjugate of the SM representation of $\Psi$.\footnote{For example, the model $Q\oplus D$ will have $Q$, $Q^c$, $D$, $D^c$, and $\phi$ as light Dq, and their GUT partners $U$, $U^c$, $E$, $E^c$, $L$ and $L^c$ as heavy Dqs.} Vectorlike mass terms in the Lagrangian will then be $M_\Psi \Psi \Psi^c$. In this way, the condensation of the dark sector does not break the electroweak symmetry \cite{VLC}. 
It is also possible to consider Dqs in the conjugate of the $SU(5)$ representations in Table \ref{tab:reps}. We denote their fragments with a tilde: if $\Psi$ is a left-handed Dq in the fundamental of $SU(\NDC)$ transforming as ${\mathcal{R}}_{\text{SM}}$, then $\widetilde{\Psi}$ is a fundamental of  $SU(\NDC)$ transforming as $ \overline{\mathcal{R}}_\text{SM}$. As above, a model containing $\widetilde{\Psi}$ contains $\widetilde{\Psi}^c$ as well. The notation is summarized in Table \ref{tab:notation}.

\begin{table}[ht!]
 \caption{Summary of the notation used in this work for the Dqs representations.
 }
    \label{tab:notation}
      \vspace{2mm}
      \renewcommand{\arraystretch}{1.3}
    \centering
    \begin{tabular}{c|c|c|c|c|c}
    \toprule
          & Poincaré & $SU(N)_{DC}$ & $SU(3)_c$ & $SU(2)_L$ & $U(1)_Y$\\
        \hline
         $\Psi$ & $(\mathbf{1/2},\mathbf{0})$ & $\square$ & $\mathcal{R}_c$ & $\mathcal{R}_L$ & $Y$ \\
         $\Psi^c$ & $(\mathbf{1/2},\mathbf{0})$ & $\overline{\square}$ & $\overline{\mathcal{R}}_c$ & $\overline{\mathcal{R}}_L$ & $-Y$ \\ 
        \hline
         $\widetilde{\Psi}$ & $(\mathbf{1/2},\mathbf{0})$ & $\square$ & $\overline{\mathcal{R}}_c$ & $\overline{\mathcal{R}}_L$ & $-Y$ \\
         $\widetilde{\Psi}^c$ & $(\mathbf{1/2},\mathbf{0})$ & $\overline{\square}$ & $\mathcal{R}_c$ & $\mathcal{R}_L$ & $Y$ \\
        \bottomrule
    \end{tabular}
\end{table}

The organization of the fermionic Dqs as in Tab. \ref{tab:reps} is to be understood as following. The models contain an (approximate, thanks to Dq masses) chiral dark flavor (DF) symmetry: $SU(N_{\text{DF}})_\Psi \otimes SU(N_{\text{DF}})_{\Psi^c}$
. Here $N_{\text{DF}}$ is half the overall number of new left-handed fermionic degrees of freedom. Turning on the weak SM interactions breaks the symmetry explicitly, so that the various flavors organize in \emph{species} belonging to definite SM representations.
At a scale $\LDC$, the dark sector confines. The Dq condensate spontaneously breaks the above chiral symmetry to the vectorial subgroup $SU(N_{\text{DF}})$, which we call DF group. As in QCD, the dark confinement produces $N_{\text{DF}}^2 -1$ pseudo-Nambu-Goldstone bosons (pNGB) in the adjoint representation of $SU(N_{\text{DF}})$ which we call interchangeably scalar dark mesons or dark pions (D$\pi$). 
The D$\pi$s may be much lighter than the other dark bound states, whose mass is of the order of $\LDC$, since they are pNGB.
The dark equivalent of the $\eta'$ is, as in QCD, expected to be heavier than the other mesons because of the axial anomaly. In particular in the model containing only $N$ and the scalar, because the full DF group is anomalous, there are no light pNGB as the ${N^c} N$ state plays the role of the $\eta'$. Bound states of two Dqs containing $\phi$ are not pNGB arising from any symmetry breaking pattern, and thus are not expected to be much lighter than the other dark states.\footnote{In this sense these states are not etymologically \emph{mesons}, and perhaps should be called \emph{baryons} (this time-honored nomenclature is even worse if one considers that the $\tau$ \emph{lepton} is heavier than some of the baryons). However in the literature the term meson is used to identify bound states of two quarks regardlessly of wether they are pNGB -- e.g. the $\rho$ particles are usually called vector-mesons, even if they are as heavy as baryons. We abide to this convention.}

\begin{table}[!tb]
  \centering
    \renewcommand{\arraystretch}{1.3}
     \caption{Allowed Yukawa terms involving the scalar Dq $\phi$ and the Higgs. Lowercase letters are the left-handed SM fields in a standard notation. Note that for each Higgs Yukawa term there is an analogous term with $\Psi \leftrightarrow \Psi^{c}/\widetilde{\Psi}$ and  $H\leftrightarrow H^{\dagger}$ (this is not true for $\phi$ Yukawa terms). Yukawa terms with tilded Dq $\tilde{\Psi}$  can be obtained starting from those in the table with the substitutions $\Psi^c \rightarrow \tilde{\Psi}$, $\phi\rightarrow \phi^{\dagger}$}
    \begin{tabular}{|c|c|c|}
    \hline
    Dq&$\phi$ Yukawa & Higgs Yukawa\\
    \hline
    \rowcolor{gray!25} $N$&none&$L^c {H^\dagger}N$, $N^{c} H ^\dagger\tilde{L}$\\
    $D$&$D^c\phi d^c$&$\widetilde{Q}HD^c$\\
    \rowcolor{gray!25}$L$&$L^c\phi l$&$L^{c}H\widetilde{E}$, $N^{c}HL$, $V^{c}HL$\\
    $U$&$U^c\phi u^c$&$\widetilde{Q}{H^\dagger}U^c$\\
    \rowcolor{gray!25}$E$&$E^c\phi e^c$&$LH^{\dagger}E^c$\\
    $Q$&$Q^c\phi q$&$Q^{c}H\widetilde{D}$, $Q^{c}{H^\dagger}\widetilde{U}$\\
    \rowcolor{gray!25}$V$&none&$V^{c}HL$\\
    any\,$\Psi$ & $\phi \Psi \widetilde{\Psi}$\quad {if}\quad $\NDC=3$& \textemdash \\
    \hline
    \end{tabular}
    \label{tab:yukawa}
\end{table}

The most important terms for both model building and phenomenology are those connecting the dark sector with the SM, namely
\be\label{eq:lag}
\lag_{\text{DS-SM}} = - \lambda_{\phi H} \phi^\dagger \phi \, H^\dagger H + \lag_{\text{Dark Yukawa}}\,.
\ee
As for the SM, the Yukawa terms are the largest source of breaking of the global symmetries. Thus, before describing the content of $ \lag_{\text{Dark Yukawa}}$, let us discuss what are the accidental symmetries enjoyed by the theory in their absence.
\begin{description}
    \item[Species Number] An independent $U(1)$ symmetry for each species of Dq as in Table \ref{tab:reps} (including $\phi$). This symmetry leads to the stability of the lightest dark mesons made of different species of Dqs ${\Psi^c}_i \Psi_j$ or ${\Psi^c}_i \phi$ ($i$ and $j$ are species indices), which are the equivalent of charged QCD pions.
    \item[G-Parity] Equivalently to QCD, if Dqs belong to non-trivial weak isospin representations with vanishing hypercharges, the theory is invariant under the discrete symmetry acting on Dqs as $\Psi \to \exp\left\{i \pi \frac{\sigma^2}{2}\right\} \Psi^c$, and trivially on the SM fields. Consequently, the lightest G-odd dark state is kept stable.
    \item[Dark Baryon Number] A subgroup of the species number symmetry under which all the Dqs rotate with the same phase. This symmetry is responsible for the stability of the lightest dark baryon (DB), i.e. the lightest composite state of $\NDC$ Dqs in a totally antisymmetric DC combination. This is equivalent to the SM baryon number symmetry, which is responsible for the stability of the proton.
\end{description}

Species and G-parity may be broken either at renormalizable level by Yukawa interactions (which generally preserve the DB symmetry) or at the level of dimension five or higher. The Yukawa terms in $\lag_{\text{Dark Yukawa}}$ are essentially of two kinds, as shown in Table \ref{tab:yukawa}. The first involves the SM Higgs field and two dark fermions: there are only a few possibilities, which lead in \cite{CDM} to a relatively small number of viable models. This changes drastically in the presence of the scalar Dq $\phi$: for any fermionic Dq with a SM counterpart, one can write a term involving them and the scalar Dqs, as in Table \ref{tab:yukawa}. These terms allow to break the species number in almost all the relevant cases as we shall see. Interestingly, thanks to them, the Dqs $Q$, $D$, and $U$ inherit the SM baryon number. In the presence of Dqs $L$ and $E$, since only one family of Dqs is introduced coupling to all three SM families, the three SM lepton numbers are broken to a single lepton number: $U(1)^3_\ell \to U(1)_\ell$; $L$ and $E$ acquire a charge under this symmetry. Similar considerations hold for the tilded versions of the mentioned Dqs. The quantum numbers are summarized in Table \ref{tab:accidental-numbers-recap}. Other light Dqs may then inherit the SM numbers through other interactions, e.g. if both $N$ and $L$ are light, $N$ acquires SM lepton number $1$ via the term $HLN^c$. In models in which $N$ is light but $L$ is not, as we shall discuss later on, the lepton flavor symmetry is preserved up to dimension five operators, at the level of which $N$ acquires lepton number $1$.

In the absence of the dark scalar, the DB symmetry is more robust than the other symmetries, since it may be broken only at dimension six or higher. An estimate of the lifetime of DBs is then
\begin{equation}\label{eq:lifetime}
    \tau_{\text{DB}} \sim \frac{8\pi \Lambda_{\slashed{\text{DB}}}^{4}}{ M_{\text{DM}}^{5}} \sim 10^{26} \si{\second} \, \left( \frac{\Lambda_{\slashed{\text{DB}}}}{M_{\text{Pl}}}\right)^{4} \ \left(\frac{100 \text{ TeV}}{M_{\text{DM}}}\right)^{5} 
\end{equation}
If we take the scale suppressing these operators to be around the Planck scale, this lifetime evades the bounds from indirect searches $\tau > 10^{26\div28}\si{\second}$ \cite{id-cohen,Aartsen_2018, HAWC:2023bti, Song:fermilat} if the mass of the DM is in the ballpark of $\SI{100}{\tera\electronvolt}$ or below. As we shall see in Sec. \ref{sec:cartoons}, this value is also selected to reproduce the correct thermal relic abundance in several scenarios.

Including a scalar Dq may introduce dimension five operators that break the DB number, depending on the specific model, such as
\be
L {H^\dagger} E \phi \quad \text{or} \quad Q {H^\dagger} D \phi\quad \text{for}\; \NDC=3\,.
\ee

If the only stable symmetry is the DB number, the lightest DB can act as the DM. On the contrary, all charged dark mesons, which arise in general and whose relic abundance is heavily constrained by observations, decay immediately. In some models species numbers are only broken at dimension five, in which case the mesons are meta-stable. If they are charged, they must decay soon enough in the cosmological history.

\begin{table}[!h]
 \caption{Quantum numbers of the Dqs under the stable DB number and the SM accidental symmetries. Tilded Dqs have opposite numbers since their SM representation is conjugated. The Dq $N$ only inherits lepton number at the level of dimension five unless $L$ is also light.}
    \label{tab:accidental-numbers-recap}
    \vspace{2mm}
    \centering
    $
    \begin{array}{|c|c|c|c|c|c|c|c|}
    \hline
     & D&U&Q&L&E&N&\phi \\
     \hline
     \rowcolor{gray!25} U(1)_{\text{DB}} & 1 &1& 1&1&1&1&1\\
    U(1)_b & -\frac{1}{3} & -\frac{1}{3} & \frac{1}{3} & 0 & 0 & 0 &0 \\
    \rowcolor{gray!25} U(1)_\ell & 0 & 0& 0&1 & -1 & 1^\ast & 0 \\
     \hline
    \end{array}
    $
\end{table}

\subsection{Hierarchy of Mass Scales } \label{sec:cartoons}

In this section, we discuss the various mass scales of our models, and how their ordering is relevant for identifying general aspects of accidental composite dark matter models. There are four basic scenarios, depicted in Fig. \ref{fig:mass-ordering}.
\input{mass-hiearchy}

There are two scales which are determined dynamically, namely the scale of confinement of the dark sector $\LDC$, and the scale of grand unification $\MGUT$. The value of the former is selected by cosmology as we shall describe shortly, depending on its relative value with respect to the masses of the Dqs, which is also crucial in understanding the spectrum of the bound states. The latter scale is determined by the unification of the SM couplings, which in turn depends on the masses $M_H$ of the heavy fermions, assumed to lie between $\MGUT$ and $\LDC$. The scale $M_H$ is not a dynamical scale, and it is represented as a range in Fig. \ref{fig:mass-ordering} as a consequence of our approach to grand unification, which is described in Sec. \ref{sec:relaxed-gut}. There are two more scales that are free parameters of the model, whose position with respect to the others determines the qualitative behaviour of the models: the mass of the scalar $M_\phi$ and the mass of the light fermions $M_\Psi$. Let us discuss the scenarios in Fig. \ref{fig:mass-ordering} from left to right.
 
If both $M_\phi$ and $M_\Psi$ lie beneath the confinement scale, the bound states are Coulomb-like. The DB have masses of the order of $\NDC \, \LDC$, while the masses of dark mesons can be estimated as described in the next section. As we shall argue later, the dark matter candidate (DMC) of our models is the lightest of the DBs. If the DM relic abundance is determined by a geometric cross section
\be
\langle \sigma v_{\text{rel.}} \rangle \sim \frac{\pi}{\LDC^2}\,,
\ee
one finds that the observed relic abundance of DM \cite{Planck:2018} is reproduced with a confinement scale in the ballpark of $\SI{100}{\tera\electronvolt}$ .\footnote{Smaller values of $\LDC$ are allowed if one assumes a matter-antimatter asymmetry in the dark sector \cite{costa:asymmetric-dm, Barr:asymmetric-dm}. In this case, the cosmological implications of the formation of dark nuclei must be taken into account \cite{Redi:dark-nuclei, Mahbubani:dark-nucleosynthesis}.} Recall that this is the same range selected from comparing the naively estimated lifetime of DBs with the bounds from indirect searches, see Eq. \eqref{eq:lifetime}. There is little to say about the hierarchy between the bound states made of only fermions (or only scalars) and the \emph{hybrid} ones made of both fermions and scalars, and thus about the nature of the DMC, without resorting to lattice calculations or other numerical means.

If both  $M_\phi$ and $M_\Psi$ lie above the confinement scale \cite{WCD}, on the other hand, the bound states are Coulomb-like, and their hierarchy (including whether the hybrid states are lighter than the purely fermionic ones) depends on the precise orderings of the Dq masses: dark hadrons made of heavier Dqs will be heavier. We shall refer to this configuration as weakly coupled scenario. The lightest dark states would be glueballs with mass $7 \LDC$ \cite{Morningstar_1999}. They cannot comprise the DM, lest overclosing the universe, thus must decay before BBN. Higher up in mass there would be dark mesons of mass $\sim 2 M_{\Psi,\Phi}$, and finally DB with mass $\sim \NDC M_{\Psi,\Phi}$. The DMC would be the DB with the lightest constituents and lowest spin. 
As described in \cite{WCD}, the dynamics of the DM freeze out and the lifetime of the dark glueballs lead to values of $\LDC$ lower than in the previous case.

The third scenario, in which the mass of the scalar lies above or in the ballpark of $\LDC$, while the light fermions are much lighter, is the one that we shall consider in the rest of this work unless otherwise specified. As far as fermionic Dqs are concerned this is another case of strongly coupled scenario, like the first. In this case, however, there is a clear hierarchy between the heavier hybrid bound states and the lighter bound states with fermionic Dqs as constituents. The latter annihilate with geometrical cross-section as in the first scenario, so cosmology once again selects $\LDC \sim \SI{100}{\tera\electronvolt}$.

In the last scenario, in which the light fermions lie above $\LDC$ with the scalar being much lighter, the nature of the DM is similar to the case of \cite{strumia-scalardm}, with differences due to the existence of the fermionic Dqs. We shall not consider this scenario any further.

\begin{table}[!ht]
    \centering
    \caption{Summary of the value of the confinement scale $\LDC$ and the nature of the DMC in the four scenarios of Fig. \ref{fig:mass-ordering}. We mostly consider the third scenario in this work, marked with an asterisk.}
    \label{tab:scenario-recap}
    \vspace{2pt}
    \begin{tabular}{ccc}
    \toprule
    Scenario  & $\LDC$ & Type of DMC \\
    \midrule
    $1$     & $\sim\SI{100}{\tera\electronvolt}$& Need non-perturbative calculations \\
    $2$ &  $<\SI{100}{\tera\electronvolt}$(see \cite{WCD}) & Depends on the Dq masses\\
    \;$3^\ast$ &$\sim\SI{100}{\tera\electronvolt}$ & fermions as constituents\\
    $4$ & $\sim\SI{100}{\tera\electronvolt}$&  scalars as constituents \\
    \bottomrule
    \end{tabular}
\end{table}

\subsection{Dark Hadrons}

At energies below the scale of confinement $\LDC$, the degrees of freedom of the dark sectors will be dark hadrons. There are essentially three types, with different mass scales and properties: dark mesons, DBs, and dark glueballs.

\subsubsection{Dark Glueballs}
The lightest glueballs have mass $7 \LDC$ \cite{Morningstar_1999}, which means that in scenarios in which the Dqs are heavier than the confinement scale, they might be the lightest dark states. In order not to overclose the universe, they cannot be stable on cosmological timescales. In particular one requires that they all decay before Big Bang Nucleosynthesis (BBN), that is, before $\SI{1}{\second}$ after the big bang \cite{BBN-glueballs, gammaray-glueballs}. We shall not consider dark glueballs any further, and we refer to \cite{WCD, Dondi:2020} for details on their impact on cosmology and collider phenomenology.

\subsubsection{Dark Baryons}

DBs are bound states of $\NDC$ Dqs in an antisymmetric DC combination. As discussed above, in strongly coupled scenarios their mass are given by the dimensional transmutation scale $M_{\text{DB}} \sim \NDC \, \LDC$, while in weakly coupled scenarios they are determined by their constituents $M_{\text{DB}} \sim \NDC \, M_{\Psi,\phi}$. Let us restrict, for now, to DBs with only fermionic Dqs as constituents. By addition of the constituents' spins, the DBs are fermions for odd values of $\NDC$, and bosons for even values. Their SM quantum numbers are determined by their DF representation and its decomposition under the SM gauge group.

In this work we seek the DM candidate (DMC) among lightest DBs, which are accidentally stable. We make the assumption that the lightest multiplet of DBs is the one with the lowest spin and vanishing orbital angular momentum of the constituents. Since the DC wave function is anti-symmetric, by Fermi's statistics this means that its representation must be symmetric in spin and DF, meaning that their spin and DF representations have the same Young tableau. For the lowest values of $\NDC$, the smallest Young tableaux for both spin and DF are
    \begin{equation}\label{eq:ytabs}
    \begin{cases}
    \yng(2,1) & \text{for } \NDC=3 \\
    \yng(2,2) & \text{for } \NDC=4\\
    \yng(3,2) &\text{for } \NDC=5 \\
    \end{cases}
    \,,
    \end{equation}
    meaning that they are either spin $\frac{1}{2}$ or spin $0$, if $\NDC$ is odd or even, respectively. 
    
    DB masses receive contributions from both the masses of their constituents and the SM gauge interactions. As discussed in Sec. \ref{sec:cartoons}, the former is particularly relevant in the weakly coupled scenario in which $\LDC \ll M_\Psi$, since the lightest DBs will be those with lightest constituents as these contributions are even larger than the spin-spin splittings. Upon the spontaneous breakdown of the electroweak symmetry, the mass of charged DBs within a multiplet is lifted by electromagnetism by $\sim \SI{166}{\mega\electronvolt}$ \cite{MDM}, which means that the lightest DB within the (lightest) multiplet is the one with the smallest SM charge and may be a viable DMC.
    
There are two notable exceptions to the above criterion, namely cases in which the DMC:
\begin{itemize}
    \item {\bf{is $N^\NDC$}}. In this case flavor cannot be anti-symmetrized, and therefore the spin must be totally symmetric and equal to $\frac{\NDC}{2}\ge\frac{3}{2}$, which is not the lowest possible (see Sec. \ref{sec:Nmodels}).
    \item {\bf{contains dark scalars.}} In this case the spin of the DMC depends on the number of scalars. We briefly discuss this possibility in Sec. \ref{sec:hybrid}.
\end{itemize}

\subsubsection{Dark Mesons}
Dark mesons are bound states of a Dq and an anti-Dq. If the constituents are both fermions, they are bosons, most interestingly vectors (dark $\rho$s) or scalars (D$\pi$s), transforming in the adjoint representation of the DF group $SU(N_{\text{DF}})$. If one of the constituents is a scalar, they are fermionic in nature and have the quantum numbers of the $SU(5)$ fragments of Tab. \ref{tab:reps}, meaning that they mix with SM fermions, in such a way that the asymptotic states are an admixture of elementary SM fermions and composite dark states.
Finally there is a scalar SM singlet $\mathcal{S} \sim{\phi^\dagger}\phi/\frac{\LDC}{4\pi}$, which mixes with the Higgs through the portal of Eq. \eqref{eq:lag}. We neglect this effect as it is small except for very large values of $\lambda_{\phi H}$ which are anyways excluded by direct searches (see Sec. \ref{sec:dd}). D$\pi$s are the pNGB associated with the spontaneous breakdown of dark chiral symmetries as described in the previous section and, as such, are expected to be much lighter than the confinement scale.\footnote{Recall that the model with only $N$ is an exception since the only meson receives mass contribution from the chiral anomaly.} The same cannot be said for mesons with $\phi$ as a constituent (be they fermions or bosons), whose mass is therefore expected to be of the order of $\LDC$ in strongly coupled scenarios.

The D$\pi$ masses receive a contribution from the SM interactions (if they are charged) arising from loops in the low energy effective field theory (EFT), and one from the constituent masses, which can be estimated via standard chiral perturbation theory techniques:
\be\label{eq:pion-mass}
\Delta_{\text{SM}} m^2_{\pi_D} \sim \left(\frac{g_{\text{SM}}}{4 \pi} \LDC\right)^2 \quad \text{and} \quad \Delta_{\text{mass}} m^2_{\pi_D} \sim M_\Psi \LDC\,,
\ee
In strongly coupled scenarios these contributions are much smaller than $\LDC$: the gauge contribution alone gives, for charged D$\pi$, $m_{\pi_D} \sim 0.1 \LDC \sim \SI{10}{\tera\electronvolt}$. As opposed to this, fermionic mesons as well as $\mathcal{S}$ and mesons containing $N$ have masses of the order of $\LDC$ as any other dark state.
In weakly coupled scenarios the masses of dark mesons are determined mostly by the constituent masses (schematically: $m_{\pi_D}\sim 2 M_{\Psi,\phi}$) and receive smaller contributions from the SM gauge interactions.

As discussed above, most models feature charged mesons, which can be at most meta-stable, that is to say, they must decay through dimension four or five operators. In the latter case, requiring that they decay before BBN poses a lower limit on their mass:
\be
\tau_{\text{dim. 5}} \sim \frac{8 \pi \Lambda^2_{\text{UV}}}{m^3_{\pi_D}} \lesssim \SI{1}{\second} \implies m_{\pi_D} \gtrsim \SI{1}{\tera\electronvolt}\left(\frac{\Lambda_{\text{UV}}}{\SI{e16}{\giga\electronvolt}}\right)^{\frac{2}{3}}\,.
\ee


\subsection{Model Selection}\label{sec:selection}

Models are considered viable if they comply with the following requests:
\begin{itemize}
    \item The DC interactions exhibits asymptotic freedom, i.e. its $\beta$-function is negative. This poses a lower bound on the number of DFs depending on $\NDC$, which is easily verified.
    \item The contribution of light Dqs to the running of the SM gauge couplings does not produce Landau poles below the Planck scale. This poses an upper bound on $\NDC$ which is typically the strongest limitation to the model selection.
    \item The model has a viable DMC, namely a dark hadron that is stable thanks to an accidental symmetry, which:
    \begin{itemize}
        \item is electrically neutral;
        \item has vanishing hypercharge 
        \cite{MDM};
        \item is uncolored%
        %
        %
        \footnote{Note that this request may be challenged if one takes into account the QCD confinement of the colored DBs. Such a scenario was considered in \cite{coloredDM}. We do not consider such a possibility in this work.}%
        ;
    \end{itemize}  
\end{itemize}
As a consequence of the first two requirements, it must have integer isospin.
Let us discuss these requirements in detail.

\subsubsection{Asymptotic Freedom}

The first term in the $\beta$-function of the $SU(\NDC)$ coupling with $N_{\text{DF}}$ flavors of fermionic Dqs and one dark scalar $\phi$ is
\be
b^0_{\text{DC}} = -\frac{11}{3} \NDC + \frac{2}{3} N_{\text{DF}} + \frac{1}{6}\,.
\ee
Requiring that the dark sector exhibits asymptotic freedom is equivalent to requiring $b^0_{\text{DC}} < 0$. This only excludes $\NDC = 3$ for more than sixteen light DFs, and we found no viable model with more than fifteen, cf. Tab. \ref{tab:all-models}. 

\subsubsection{Running of the Standard Model Gauge Couplings}

The contribution of the Dqs to the SM $\beta$-functions are summarized in the rightmost columns of Table \ref{tab:reps}. Since each Dq is a left-handed fundamental of $SU(\NDC)$ and vector-like with respect to the SM, the numbers in that columns are to be multiplied by $2 \NDC$. Assuming that in strongly coupled scenarios the Dqs start contributing to the running when the scale reaches $\LDC \sim \SI{100}{\tera \electronvolt}$, one finds the following conditions \cite{CDM}:
\be
\NDC \sum_\Psi \Delta b^\Psi_Y \lesssim \frac{11}{2} \quad \NDC \sum_\Psi \Delta b^\Psi_2 \lesssim 5 \quad \NDC \sum_\Psi \Delta b^\Psi_3 \lesssim 5\,,
\ee
where the sum extends over all the light species of Dqs. The Dqs $T$, $S$, $G$, and $X$ are automatically excluded since they contribute too much to the running.

In weakly coupled scenarios the contributions of Dqs to the running may start at higher scales, resulting in weaker restrictions on the content of the model and one must therefore check the perturbativity of the SM gauge couplings case by case.

\subsubsection{Dark Matter Candidate}

A dark state is stable thanks to an accidental symmetry if it is the lightest particle charged under that symmetry. If, as we are requiring, all global symmetry are broken (at most at the level of dimension five) except for the DB symmetry, the DMC must be sought among the lightest DBs.

In strongly coupled scenarios $m_\Psi\ll\LDC$, these are simply those of Eq. \eqref{eq:ytabs}. In order for one of those DBs to be a viable candidate, a color singlet with zero hypercharge must be found in the decomposition of the DF representation under the SM gauge group. The model must be discarded if this is not the case, as it will have stable particles in the spectrum that do not evade observational bounds. In Appendix \ref{app:examples} we furnish a few examples of both models that exhibit such a state and models that do not.

If on the other hand, all the Dqs are heavier than $\LDC$, we assume that the lightest DB is that of lowest spin among those containing the lightest Dqs as constituents. Indeed in this case the splittings due to spin-spin interactions are expected to be smaller than in the previous case and the mass orderings of the fundamental Dqs dictates the hierarchy between the bound states.

\subsection{Hybrid Candidates}\label{sec:hybrid}
\input{composite-dm/hybrid-dmc}

%% file: composite-dm/mass-hiearchy.tex
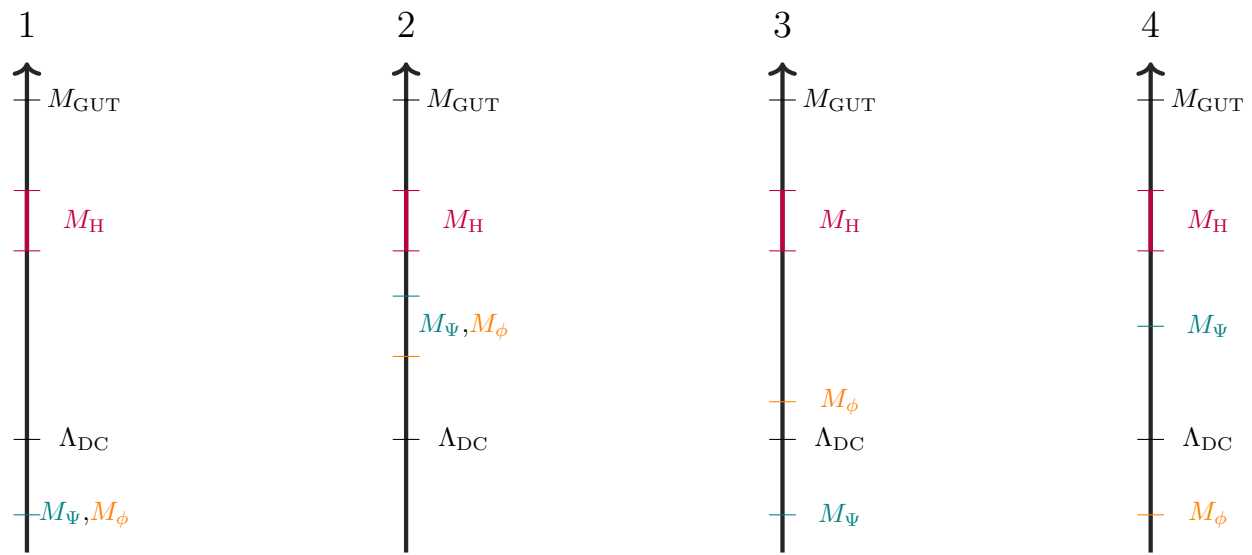
\begin{figure}[!bh]
\centering
\begin{tikzpicture}
\coordinate (lambdaDC) at ($(0,3.5cm)$) {};
\coordinate (MHlo) at ($(0,6cm)$) {};
\coordinate (MHmid) at ($(0,6.4cm)$) {};
\coordinate (MHhi) at ($(0,6.8cm)$) {};
\coordinate (MGUT) at ($(0,8cm)$) {};
\coordinate (ML) at ($(0,2.5 cm)$) {};
\node at ($(lambdaDC)+(5ex,0)$) {$\Lambda_{\text{DC}}$};
\node at ($(MHmid)+(5ex,0)$) {\color{purple}$M_\text{H}$};
\node at ($(MGUT)+(5ex,0)$) {$M_{\text{GUT}}$};
\node at ($(MGUT)+(0,1cm)$) {\Large$1$};
\node at ($(ML)+(5ex,0)$) {{\color{teal}$M_\Psi$},{\color{orange}{$M_{\phi}$}}};
\draw [ultra thick,  ->, black!85] (0,2cm) -| ($(MGUT)+(0,0.5cm)$);
\draw ($(lambdaDC)+(5pt,0)$) -- ($(lambdaDC)-(5pt,0)$);
\draw[purple] ($(MHlo)+(5pt,0)$) -- ($(MHlo)-(5pt,0)$);
\draw[purple] ($(MHhi)+(5pt,0)$) -- ($(MHhi)-(5pt,0)$);
\draw ($(MGUT)+(5pt,0)$) -- ($(MGUT)-(5pt,0)$);
\draw [teal] ($(ML)+(5pt,0)$) -- ($(ML)-(5pt,0)$);

\draw [ultra thick, purple] (MHlo) -| (MHhi);
\end{tikzpicture}
\hfill
\begin{tikzpicture}
\coordinate (lambdaDC) at ($(0,3.5cm)$) {};
\coordinate (MHlo) at ($(0,6cm)$) {};
\coordinate (MHmid) at ($(0,6.4cm)$) {};
\coordinate (MHhi) at ($(0,6.8cm)$) {};
\coordinate (MGUT) at ($(0,8cm)$) {};
\coordinate (MLmid) at ($(0,5 cm)$) {};
\coordinate (MLhi) at ($(0,5.4 cm)$) {};
\coordinate (MLlo) at ($(0,4.6 cm)$) {};
\node at ($(lambdaDC)+(5ex,0)$) {$\Lambda_{\text{DC}}$};
\node at ($(MHmid)+(5ex,0)$) {\color{purple}$M_\text{H}$};
\node at ($(MGUT)+(5ex,0)$) {$M_{\text{GUT}}$};
\node at ($(MGUT)+(0,1cm)$) {\Large$2$};
\node at ($(MLmid)+(5ex,0)$) {{\color{teal}$M_\Psi$},{\color{orange}{$M_{\phi}$}}};
\draw[ultra thick,  ->, black!85] (0,2cm) -| ($(MGUT)+(0,0.5cm)$);
\draw ($(lambdaDC)+(5pt,0)$) -- ($(lambdaDC)-(5pt,0)$);
\draw[purple] ($(MHlo)+(5pt,0)$) -- ($(MHlo)-(5pt,0)$);
\draw[purple] ($(MHhi)+(5pt,0)$) -- ($(MHhi)-(5pt,0)$);
\draw[orange] ($(MLlo)+(5pt,0)$) -- ($(MLlo)-(5pt,0)$);
\draw[teal] ($(MLhi)+(5pt,0)$) -- ($(MLhi)-(5pt,0)$);
\draw ($(MGUT)+(5pt,0)$) -- ($(MGUT)-(5pt,0)$);

\draw[ultra thick, purple] (MHlo) -| (MHhi);
\path [bottom color=orange, top color=teal, shaded path={ 
    [draw=transparent!0, ultra thick] (MLlo) -- (MLhi)
}];
\end{tikzpicture}
\hfill
\begin{tikzpicture}
\coordinate (lambdaDC) at ($(0,3.5cm)$) {};
\coordinate (MHlo) at ($(0,6cm)$) {};
\coordinate (MHmid) at ($(0,6.4cm)$) {};
\coordinate (MHhi) at ($(0,6.8cm)$) {};
\coordinate (MGUT) at ($(0,8cm)$) {};
\coordinate (Mphi) at ($(0,4 cm)$) {};
\coordinate (ML) at ($(0,2.5 cm)$) {};
\node at ($(lambdaDC)+(5ex,0)$) {$\Lambda_{\text{DC}}$};
\node at ($(MHmid)+(5ex,0)$) {\color{purple}$M_\text{H}$};
\node at ($(MGUT)+(5ex,0)$) {$M_{\text{GUT}}$};
\node at ($(MGUT)+(0,1cm)$) {\Large$3$};
\node at ($(ML)+(5ex,0)$) {\color{teal}$M_\Psi$};
\node at ($(Mphi)+(5ex,0)$) {\color{orange}$M_{\phi}$};
\draw [ultra thick,  ->, black!85] (0,2cm) -| ($(MGUT)+(0,0.5cm)$);
\draw ($(lambdaDC)+(5pt,0)$) -- ($(lambdaDC)-(5pt,0)$);
\draw[purple] ($(MHlo)+(5pt,0)$) -- ($(MHlo)-(5pt,0)$);
\draw[purple] ($(MHhi)+(5pt,0)$) -- ($(MHhi)-(5pt,0)$);
\draw ($(MGUT)+(5pt,0)$) -- ($(MGUT)-(5pt,0)$);
\draw [teal] ($(ML)+(5pt,0)$) -- ($(ML)-(5pt,0)$);
\draw [orange] ($(Mphi)+(5pt,0)$) -- ($(Mphi)-(5pt,0)$);

\draw [ultra thick, purple] (MHlo) -| (MHhi);
\end{tikzpicture}
\hfill
\begin{tikzpicture}
\coordinate (lambdaDC) at ($(0,3.5cm)$) {};
\coordinate (MHlo) at ($(0,6cm)$) {};
\coordinate (MHmid) at ($(0,6.4cm)$) {};
\coordinate (MHhi) at ($(0,6.8cm)$) {};
\coordinate (MGUT) at ($(0,8cm)$) {};
\coordinate (ML) at ($(0,5 cm)$) {};
\coordinate (Mphi) at ($(0,2.5cm)$) {};
\node at ($(lambdaDC)+(5ex,0)$) {$\Lambda_{\text{DC}}$};
\node at ($(MHmid)+(5ex,0)$) {\color{purple}$M_\text{H}$};
\node at ($(MGUT)+(5ex,0)$) {$M_{\text{GUT}}$};
\node at ($(MGUT)+(0,1cm)$) {\Large$4$};
\node at ($(ML)+(5ex,0)$) {\color{teal}$M_\Psi$};
\node at ($(Mphi)+(5ex,0)$) {\color{orange}$M_{\phi}$};
\draw [ultra thick,  ->, black!85] (0,2cm) -| ($(MGUT)+(0,0.5cm)$);
\draw ($(lambdaDC)+(5pt,0)$) -- ($(lambdaDC)-(5pt,0)$);
\draw[purple] ($(MHlo)+(5pt,0)$) -- ($(MHlo)-(5pt,0)$);
\draw[purple] ($(MHhi)+(5pt,0)$) -- ($(MHhi)-(5pt,0)$);
\draw ($(MGUT)+(5pt,0)$) -- ($(MGUT)-(5pt,0)$);
\draw [orange] ($(Mphi)+(5pt,0)$) -- ($(Mphi)-(5pt,0)$);
\draw [teal] ($(ML)+(5pt,0)$) -- ($(ML)-(5pt,0)$);

\draw [ultra thick, purple] (MHlo) -| (MHhi);
\end{tikzpicture}
\hfill
\caption{Cartoons with the four scenarios for the mass orderings in our models. Here $M_{\text{GUT}}$ is the scale of grand unification of the SM couplings, $\LDC$ is the scale of confinement of the dark sector, $M_\phi$ is the mass of the scalar Dq, while $M_\Psi$ and $M_H$ are the mass scales of the light and heavy fermion Dqs, respectively. The latter is presented as a range, since, as discussed in Sec. \ref{sec:GUT}, our relaxed criterion for the unification does not fix it univocally as a function of the other mass scales.}
\label{fig:mass-ordering}
\end{figure}

%% file: composite-dm/hybrid-dmc.tex
To the best of our knowledge, there is no literature regarding the dynamics of the formation of hybrid bound states with both fermion and scalar constituents in confining theories, which may for instance be attained through lattice simulations. This makes it challenging to address the question of the mass hierarchy between the bound states, which is needed for instance to assess the nature of the DMC in a specific model in the strongly coupled scenario. In QCD, hybrid hadrons are those with both quarks and gluons as valence constituents. There is plenty of literature on hybrid mesons, but hybrid baryons have attracted less attention for phenomenological reasons \cite{qcd-hybrids}. Lattice simulations seem to indicate that baryons with a gluonic component lie slightly heavier than the lightest ordinary baryons \cite{qcd-hybrid-baryons}.

We would like to have an argument to understand the hierarchy between hybrid and regular states in our DM models. In the third and fourth scenarios of Sec. \ref{sec:cartoons} the hierarchy of the DBs is dictated by the hierarchy between the scalar and fermionic Dqs. In the other two scenarios if the masses of the two types of constituents are comparable, one may argue once more that the DMC will be the DB with the smallest spin. If, then, a hybrid DB has smaller spin than the lightest DB with only fermionic or only scalar constituents, it may be the DMC. 
A rough prescription for understanding whether this is the case is the following. 

Let a hybrid DB contain $N_F$ fermionic Dqs and $N_S$ scalars, with $\NDC = N_F + N_S$, and let us temporarily disregard whether the dark fermions are in a flavor representation that contains a viable DMC. We treat the fermions and the scalars separately, establishing their total angular momenta $J_F$ and $J_S$. The spin of the lightest hybrid state with $N_F$ fermions and $N_S$ scalars will then be the smallest representation arising from the combination of $J_F$ and $J_S$. One then in principle compares this value with the spin of the lightest DB with only fermionic or only scalar constituents. Establishing what is the spin of the latter is challenging and beyond the scope of this work, which unfortunately means that we are not able to give a definitive answer to the question.

The fermions will arrange in the spin configurations arising from the composition of $N_F$ spin $\frac{1}{2}$. Indeed the total antisymmetry of the wavefunction is enforced by taking a flavor representation whose symmetry matches that of the spin representation. If $N_F$ is odd, $J_F$ always starts from $\frac{1}{2}$, while if $N_F$ is even, $J_F=0, 1$ are always possible. An exception is the case in which the Dq $N$ is the lightest fermionic Dq, with $M_N \sim M_\phi$, in which case the spin configurations is necessarily $\frac{N_F}{2}$ (see Eq. \eqref{eq:ytab-Nmodels}; these models are only viable in the weakly coupled scenario). The scalars, on the other hand, have to arrange in a totally antisymmetric configuration in order to respect the correct statistics, and since they enjoy no DF symmetry, 
orbital angular momenta of the constituents have to be introduced.

Let us consider the two easiest cases of $N_S=1,2$, and first look at the case in which $N$ is not among the lightest fermionic Dqs. If there is only one scalar, it will clearly contribute with $J_S=0$. If there are two scalars, in order to respect the bosonic symmetry they must have an odd angular momentum, $J_S=1,3,5,\ldots$, since an eigenstate of the total angular momentum has parity $\left(-1\right)^J$  under the exchange of the two particles. In light of what we observed above on $J_F$, we see that in these two easiest cases the smallest value arising from the combination of $J_S$ and $J_F$ is 
\be
J_{N_F+N_S}=\left\{
\begin{aligned}
    \frac{1}{2} \quad &\text{for odd}\;N_F  \\
    0 \quad &\text{for even}\;N_F 
\end{aligned}
\right. 
\ee
Recall that DBs with fermionic constituents have $J=\frac{1}{2}$ for odd $\NDC$ and $J=0$ for even $\NDC$, which means that if $N_S=1$ and $N_F$ is even, the hybrid DBs have lower spin than the purely fermionic DBs, and may be the DMC. If $N_F=2$ the only possible viable hybrid DMC would be $\Psi \widetilde{\Psi} \phi$. However in models containing $\Psi \oplus \widetilde{\Psi}$ for $\NDC=3$ the DB number is broken explicitly by the last interaction of Tab. \ref{tab:yukawa}, so this case must be discarded. One may then have hybrid DMC for $N_S = 1$, $\NDC=5, 7$.

If, instead, $N$ is the lightest fermionic Dq, one has,
\be\label{eq:hybrid-with-N-spin}
J_{N_F+N_S}=\left\{
\begin{aligned}
    \frac{N_F}{2} \quad &\text{for}\;N_S=0, 1  \\
    \left|1-\frac{N_F}{2}\right| \quad &\text{for}\;N_S=2 \quad\text{up to $N_F=4$}
\end{aligned}
\right. 
\ee
and one sees that $N_S=2$ is always preferred with respect to $N_S=1,0$ until $\NDC=6$. For instance for $\NDC=3$ the Dirac DB $N\phi^2$ may be the DMC.

In the weakly coupled scenario, in which the fundamental constituents are non-relativistic, one may be able to deduce the total angular momentum $J_S$ of the subset of scalar Dqs inside the DB even for $N_S\ge 3$ \cite{strumia-scalardm}. We leave a more exhaustive treatment of hybrid DBs to a future work.

%% file: model-building/classification.tex
In this section we show a classification of models that are viable according to the criteria of Sec. \ref{sec:selection}. This classification is only viable in the strongly coupled scenario. Recall, indeed, that in the weakly coupled scenario, if the mass of the light Dqs is large enough, several more models may pass the selection, since the bounds arising from the perturbativity of the SM are weakened. We also give a separate discussion on certain models with light $N$ that are only viable in the weakly-coupled scenario.

\subsection{Minimal Models} \label{sec:minimal-models}

In principle, one can construct every possible model compatible with our assumptions of Sec. \ref{sec:selection} by considering all possible combinations of Dqs taken from Tab. \ref{tab:reps} and excluding those that do not meet the requirements, arriving at the list in Table \ref{tab:all-models} in App. \ref{app:models}.  Many models will present the same DMC, as there are but a few possibilities to combine the fragments of Tab. \ref{tab:reps} to form a DB that is a color singlet with vanishing hypercharge. In particular, if fragments from more than three different $SU(5)$ representations are considered, one either encounters subplanckian Landau poles or falls back to the case of a smaller model, extended with some light Dqs that do not constitute the DMC. Indeed, one can interpret the list in terms of \emph{minimal models}, namely those models with the smallest light Dq content needed to produce a given DMC, and their extensions. The minimal models with the respective DMC (from which one can read off the allowed number of DCs) are envisioned in Tab. \ref{tab:Stable_Minimal} with the exclusion of those containing $N$, which are discussed separately.

Minimal models can be extended in two ways (see App. \ref{app:extension} 
        for examples): 
\begin{enumerate}
        \item Adding any light Dq (including $N$) and keeping $\NDC$ fixed to find a model with the same DMC 
        \item Adding the Dq $N$ and increasing $\NDC$ by one to form a new minimal model whose DMC contains the $N$ Dq in addition to the previous content \footnote{There are only three viable models with $N$ that cannot be obtained directly by extension from minimal models listed in Table \ref{tab:Stable_Minimal} : for $N_{DC} = 4$ $L\oplus E \oplus N$ and $L\oplus \widetilde{L} \oplus N$   with candidate of the type respectively $LLEN$ and $ L\widetilde{L}NN$, and for $N_{DC} = 5$ the model $L\oplus E \oplus N$ with candidate $LLENN$.}
\end{enumerate}
Moreover if it is possible to rise $N_{DC}$ without spoiling any of the requirements, one can in principle realize a minimal model in which the DMC is a hybrid state containing scalars in addition to the previous content, as shown in the table. In Sec. \ref{sec:hybrid} we give a discussion on 
hybrid states, arguing that assessing if hybrid DBs may be the DMC is a difficult task. For completeness, however, we do include in both Tab. \ref{tab:Stable_Minimal} and Tab. \ref{tab:all-models} models in which the DMC is a hybrid DB. 

The requirement for the SM couplings to stay perturbative below the Plank scale implies that there are no minimal models with more than three different species of light Dqs (excluding $N$: indeed, the minimal model $N$ and its extensions are in some sense special and deserve a separate treatment). Note that no minimal model with $V$ in the light Dq spectrum is quoted. The reason for this is that in order to avoid subplanckian Landau poles in the SM, the only possibility for such models would be $\NDC=3$. However, in that case, the combination of the dimension five operators
\be
V^c \phi H l\quad \text{and} \quad V \sigma^\mu V^{c\dagger} \left(D_\mu \phi \right)
\ee
breaks the DB symmetry, rendering any model containing $V$ unacceptable. The same operators can be written with $N$ in place of $V$:
\be\label{eq:Nbreak3colors}
N^c \phi H l\quad \text{and} \quad N \sigma^\mu N^{c\dagger} \left(D_\mu \phi \right)\,,
\ee
which means that models with light $N$ are only acceptable for $\NDC\ge4$.

\begin{table}[!ht]
\renewcommand{\arraystretch}{1.2}
    \centering
    \caption{Minimal models, i.e. models with the smallest Dq content that produce a given DMC (third column). The allowed value of $\NDC$ can be read off the number of constituents of the various DMCs. In the last column we show the $SU(2)_L$ representation of the DMC (recall that it must be a color singlet with vanishing hypercharge).}
    \begin{tabular}{|c|c|c|c|}
    \hline
         $SU(5)$ & Minimal Model & DMC & $SU(2)_L$ Multiplet\\
        \hline
        \rowcolor{yellow!25} \multirow{2}*{$\fivebar \oplus \five $}&$D \oplus \widetilde{D}$& \makecell*
        {$D\widetilde{D}\phi^2$, $D\widetilde{D}\phi^3$, $D \widetilde{D} \phi^4$, $D\widetilde{D}\phi^5$\\
        $D\widetilde{D}D\widetilde{D}$, 
        $D\widetilde{D}D\widetilde{D}\phi$,  $D \widetilde{D}D\widetilde{D}\phi^2$, $D\widetilde{D}D\widetilde{D}\phi^3$, \\
        $D \widetilde{D}D\widetilde{D}D\widetilde{D}$, $D\widetilde{D}D\widetilde{D}D\widetilde{D}\phi$ 
        }
        &{\colorbox{yellow!25}{\makecell*{$\one$\\$2 \times \one$\\$2 \times \one$
        }}} \\ 
 \cline{2-4}
  \rowcolor{yellow!25} & $L \oplus \widetilde{L}$&{\colorbox{yellow!25} {\makecell*{$L\widetilde{L}\phi^2$, $L\widetilde{L}\phi^3$, $L \widetilde{L} \phi^4$, $L\widetilde{L}\phi^5$\\
        $L\widetilde{L}L\widetilde{L}$, 
        $L\widetilde{L}L\widetilde{L}\phi$,  $L \widetilde{L}L\widetilde{L}\phi^2$, $L\widetilde{L}L\widetilde{L}\phi^3$, \\
        $L\widetilde{L}L\widetilde{L}L\widetilde{L}$, $L\widetilde{L}L\widetilde{L}L\widetilde{L}\phi$}} }&  \makecell*{$\one\oplus \three$\\$2 \times \one\oplus \three\oplus \five$\\ $2 \times \one\oplus 2 \times \three \oplus \five \oplus \seven$} \\
  \cline{2-4}
  \rowcolor{yellow!25} & $D\oplus \widetilde{D}\oplus L \oplus \widetilde{L}$&{\colorbox{yellow!25} {\makecell*{$D\widetilde{D}\phi^2 $,\\$L\widetilde{L} \phi^2$,\\$D\widetilde{D}D\widetilde{D}$,\\ $L\widetilde{L}L\widetilde{L}$,\\ $D\widetilde{D}L\widetilde{L}$}} }&{\colorbox{yellow!25}{\makecell*{$\one$\\$\one\oplus\three$\\$2\times\one$\\$2\times\one\oplus \three \oplus \five$\\ $ 2\times\one\oplus  2\times\three$}}} \\
  \cline{1-4}
 \rowcolor{cyan!25} $\ten \oplus \mathbf{\overline{10}}$& $E \oplus \widetilde{E}$ &{\colorbox{cyan!25}{\makecell*{$E\widetilde{E}E\widetilde{E}$, $E\widetilde{E} \phi^2$}}} & $\one$ \\
   \cline{1-4}
    \rowcolor{orange!25}\multirow{2}*{$\fivebar \oplus \ten$}&$D \oplus U$ & {\colorbox{orange!25}{\makecell*{$DDU$, $DDU\phi$}}} &{\colorbox{orange!25}{\makecell*{$\one$
    }}}\\
   \cline{2-4}
  \rowcolor{orange!25}  &$L \oplus E$&{\colorbox{orange!25}{\makecell*{$LLE\phi$, $ LL E \phi^2$}}} & $\one\oplus \three$ \\
   \cline{1-4}
  \rowcolor{green!25} \multirow{2}*{$\five \oplus \ten$} & $Q \oplus \widetilde{D}$ &{\colorbox{green!25}{\makecell*{$QQ\widetilde{D}$, $QQ\widetilde{D}\phi$}}} & {\colorbox{green!25}{\makecell*{$\one\oplus\three$
  }}}\\
   \cline{2-4}
  \rowcolor{green!25} & $\widetilde{D} \oplus E \oplus U$ & $\widetilde{D}EU$ & $2 \times \one$ \\
   \cline{1-4}
    \end{tabular}
   
    \label{tab:Stable_Minimal}
\end{table}


\subsection{\texorpdfstring{$N$}{N} Models}\label{sec:Nmodels}

In this section we describe the \emph{$N$ models}, in which the DMC has only $N$ as fermionic constituents. 
It is easy to see that the minimal model with this DMC contains only $N$ as a light Dq, and indeed any $N$ model can be seen as an extension of it. Since $N$ is a total SM singlet, it does not contribute to the running of the couplings, and this model could exist for any value of $\NDC\ge4$ (see above), safe from subplanckian Landau poles. Hence, the allowed values for $\NDC$ in models that are its extensions are dictated by the other light Dqs.

First let us discuss the accidental symmetries of this kind of models. If the Dq $L$ is light, the species symmetry $U(1)_N \otimes U(1)_L \otimes U(1)_\phi$ is broken to the DB symmetry $U(1)_{\text{DB}}$ at renormalizable level by the Yukawa coupling with the SM Higgs field: $L H N^c$. The same can be said if $\widetilde{L}$ is light. Otherwise, the breaking takes place at dimension five through the operator $N^c \phi \widetilde{H} l$, regardless of the number of DCs. This means that, in this second case, the model will present metastable dark mesons. 
In general these states will be charged under QCD or even electromagnetic interaction, thus their abundance must be very small and they must have all decayed before BBN.
If they come to dominate the energy budget of the universe their decay will inject significant entropy in the SM potentially leading to a phase of dilution of the DM relic abundance (see \cite{Dondi:2020} and references therein). 
. Incidentally, these are the only models that we can construct that present this feature, since in every other model the species symmetries are broken at the renormalizable level, and we can assume that mesons decay immediately into SM particles. Note that the operator that breaks the species of $N$ at dimension five also breaks the SM lepton flavor symmetry to a single lepton number, under which $N$ inherits charge $1$ as shown in Tab. \ref{tab:accidental-numbers-recap}.

Let us now discuss the properties of the DMC, which can be $N^\NDC$ or any hybrid baryon composed of $N$ and $\phi$, such as $\phi N^{\NDC-1}$. In selecting viable models, we have so far assumed that the lightest DB multiplets were those of lowest spin, whose DF representation had the Young tableaux depicted in Eq. \eqref{eq:ytabs}. We concluded then that for odd (even), $\NDC$ the DM has spin $\frac{1}{2}$ (spin $0$). When only the Dq $N$ constitutes the DM, however, one cannot antisymmetrize flavor, so $N^\NDC$ belongs to a higher spin multiplet, namely:
\be\label{eq:ytab-Nmodels}
\underbrace{
\begin{ytableau}
    ~    & & \none[\dots] &      \\     
\end{ytableau}
}_\NDC \quad \text{with spin}\; \frac{\NDC}{2}\,.
\ee
Hybrid candidates may have lower spin, but in general larger than $\frac{1}{2}$ or $1$. As a consequence, unless $N$ is the only light Dq, in which case $N^\NDC$ is the only (purely fermionic) DB and as such it is stable, one has to enforce that the lightest DB does not have any fermionic Dq other than $N$ as consitutents. To do so one may require that the mass of all the Dqs be much larger than the confinement scale $\LDC$ -- which would correspond to the weakly coupled scenario, the second of Sec. \ref{sec:cartoons} -- and that $N$ is much lighter than all the others. If the mass of the scalar $\phi$ is comparable to the mass of $N$, hybrid DBs might be lighter than $N^\NDC$, as discussed in Sec. \ref{sec:hybrid}.

Thus, a model with $N$ and other light Dqs has to be considered in two separate regimes. In the case we have just described, the DMC would be either $N^\NDC$ or a hybrid baryon, which will be much lighter than any possible DB made of the other Dqs
. In any other scenario 
the DMC is to be sought among the DB made of the other Dqs, including, possibly, hybrid DB containing $\phi$.

%% file: gut/gut.tex
In this section we analyse the models found in the previous sections in the context of an $SU(5)$ grand unification scheme 
\cite{SU5GUT}. This was done in \cite{CDM} as well, finding that essentially only one of their viable models, namely $Q\oplus\widetilde{D}$, produced a successful unification. In line with \cite{unificaxion}, the authors observe that in order for a model to succeed in the unification, either the Dq $Q$ or the Dq $V$ (or both) must be light.

\subsection{Standard Approach to Unification}

The one-loop level running of the gauge couplings is given by
\be
\invalpha{i}{\mu} = \invalpha{i}{M_Z} - \frac{b^{\text{SM}}_i}{2 \pi}\log{\frac{\mu}{M_Z}} + \delta_i(\mu)\,,
\ee
where $\alpha_1 = \frac{3}{5} \alpha_Y$, the values of the coupling constants at the $Z$ boson mass $M_Z$ \cite{ParticleDataGroup:2020ssz}, and $b^{\text{SM}}_i$ are the three SM beta-function coefficients
\be
b^\SM_1 = \frac{41}{10},\quad b^\SM_2=-\frac{19}{6}, \quad b^\SM_3 = - 7 \,.
\ee
The new physics contributions are all encoded in the $\delta_i$
, the SM corresponding to $\delta_i = 0$. Only two independent combinations of the $\delta_i$ are relevant for unification, for instance $\delta_{12} = \delta_1 - \delta_2$ and $\delta_{32} = \delta_3 - \delta_2$. 

We perform this analysis in the strongly coupled scenario, and posit that the light fermionic Dqs start to contribute to the running at $\LDC\sim\SI{100}{\tera\electronvolt}$. The heavy GUT partners are assumed to start contributing from their (common, for simplicity) mass scale $M_H$, which we take to lie between $\LDC$ and the grand unification scale $\MGUT$. Then:
\be\label{eq:deltas}
\delta_i \left(\mu\right) = - 2 \NDC \frac{
\Delta b_i}{2 \pi} \log \frac{M_H}{\LDC} - 2 \NDC \frac{\Delta b_{\text{\footnotesize full}}}{2 \pi}\log \frac{\mu}{M_H}\,,
\ee
where $\Delta\,b_i$ is the sum of the contributions from all light Dqs to the $i$-th $\beta$-function as in Tab. \ref{tab:reps}, and $\Delta b_{\text{\footnotesize full}}$ is the contribution from the full $SU(5)$ multiplet above $M_H$.

Typically, one requires that the new physics make the unification of couplings exact at a certain energy scale. As a further criterion, one requires that GUT coupling is in the perturbative range $\left[0,4 \pi\right]$. This fixes $M_{\text{GUT}}$ and $M_H$. In this way, following the same steps of \cite{CDM}, we find
%
%
that the models shown in Table \ref{tbl:GUT}, obtained extending minimal models in Table \ref{tab:Stable_Minimal}, provide a successful unification.

\begin{table}[!httbp]
  \centering
  \caption{
  Models in which the unification of the SM couplings is exact. The DMC is shown, as well as the value of the GUT coupling constant, the intermediate scale of the heavy dark fermions, and the unification scale.}
  \vspace{5pt}
 {\renewcommand{\arraystretch}{1.2}
  \begin{tabular}{|c|c|c|c|c|}
    \hline
Model&DMC&$\alpha_{\text{GUT}}$&$M_{H}$ (GeV)&$\MGUT$ (GeV)\\
\hline
\rowcolor{yellow!25}$Q\oplus\widetilde{D}$&$QQ\widetilde{D}$&$6\times 10^{-2}$&$2\times 10^{11}$&$2\times 10^{17}$\\

\rowcolor{cyan!25}$Q\oplus\widetilde{D}$&$QQ\widetilde{D}\phi$&$2.29\times 10^{-1}$&$4\times 10^{9}$&$2\times10^{17}$\\

\rowcolor{orange!25}$\widetilde{Q}\oplus D\oplus U\oplus L$&{\colorbox{orange!25}{\makecell*{$\widetilde{Q}\widetilde{Q}D$\\\text{or}\\$DDU$}}}&$8.43\times 10^{-2}$&$2\times 10^{17}$&$2\times 10^{17}$\\

\rowcolor{green!25}$Q\oplus \widetilde{D}\oplus E$&$QQ\widetilde{D}$&$4.5\times 10^{-2}$&$2\times 10^{17}$&$2\times 10^{17}$\\

\rowcolor{red!25}$Q\oplus \widetilde{D}\oplus E$&$QQ\widetilde{D}\phi$&$1.13\times 10^{-1}$&$2\times 10^{14}$&$2\times 10^{17}$\\
 \hline  
  \end{tabular}
  }

  \label{tbl:GUT}
\end{table}

In the weakly coupled scenario one can perform a similar analysis, and take the light Dq masses as the scale where they start contibuting to the running. In principle many more models could be found that provide a successful unification, thanks to the interplay between the light Dq masses and $M_H$.

\subsection{Relaxed Criterion for Unification} \label{sec:relaxed-gut}

We also follow a different approach \cite{verma}. We relax the condition that the grand unification is exact, and only require that it is in some sense \emph{better} than in the SM, albeit possibly still imperfect. In particular we require that the area of the triangle formed by the intersections of the lines drawn by the running in the $\left(\log\mu,\,\alpha^{-1}\right)$ plane is smaller than in the SM. One can then take $\MGUT$ and $\alpha^{-1}_{\text{GUT}}$ to be the coordinates of the barycenter of the triangle. As we shall see momentarily, this approach allows to select a larger number of models as successful with respect to those in Tab. \ref{tbl:GUT}. As a check of its validity we verify that the results of Tab. \ref{tbl:GUT} are recovered if one requires that the area formed by the triangle vanishes, which would correspond to exact unification with lines meeting at one point.

\begin{figure}[!htt]
    \centering
    \includegraphics[width=0.45\textwidth]{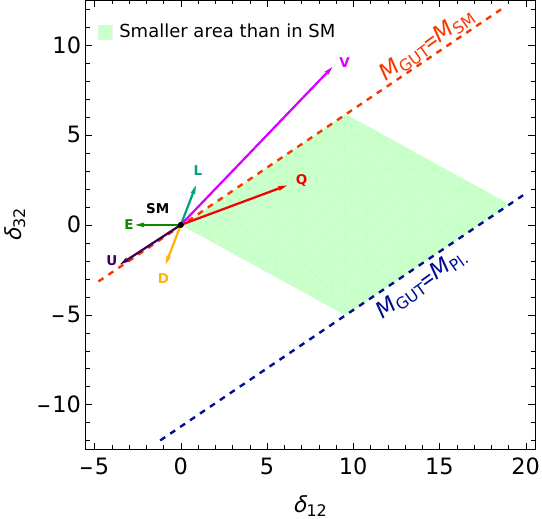}
    \hfill
    \includegraphics[width=0.45\textwidth]{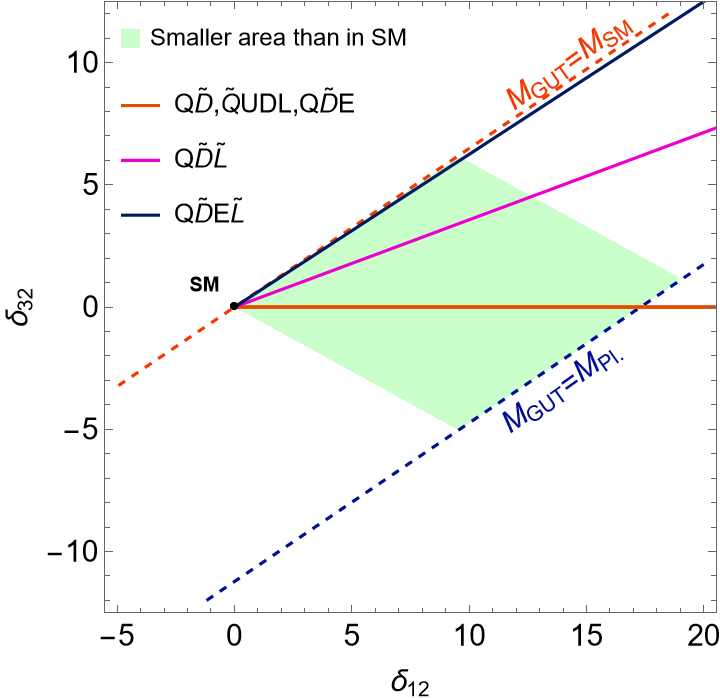}
     \caption{Modification to the grand unification of the SM couplings as a function of the contributions to the  $\beta$-functions from new physics (see main text).
     {\bf Left}: The contribution from single Dq are shown assuming $\NDC=3$, and $M_H=10^3 \LDC$. We see that only the Dqs $Q$ and $V$ point in the right direction for improving unification. {\bf Right}: Strongly coupled models with improved unification are shown as continuous half lines. The larger $M_H$, the farther one moves away from the origin along the line.}
    \label{fig:gut}
\end{figure}

In Fig. \ref{fig:gut} we summarize the results of this analysis. On the left, the contribution arising from single Dqs are shown as arrows in the $\left(\delta_{12},\,\delta_{32}\right)$ plane, whose slopes are $\frac{\Delta\,b_{32}}{\Delta\,b_{12}}$, independently of $\NDC$. The direction where the arrows point depends on the sign of $\Delta\,b_{12}$, and their length are obtained taking $M_H=10^3 \LDC$ and $\NDC=3$ as an example. The orange (blue) dashed line excludes regions in the plane in which the grand unification scale is smaller than that of the SM (larger than the Planck scale), and the shaded green region is that in which the unification is improved according to our criterion. 
The arrows stem from the origin, where the SM ($\delta_i = 0$) lies. According to Eq. \eqref{eq:deltas}, any model lies in the origin if $M_H=\LDC$: indeed in such case the lines in the $\left(\log\mu,\,\alpha^{-1}\right)$ plane (and thus the triangle they form) are just moved rigidly along the $\alpha^{-1}$ axis, and the area equals that of the standard case. By inspecting the arrows we see that relaxing the criterion for unification does not change the conclusion that, in order for a model to provide an acceptable unification, either the Dq $Q$ or the Dq $V$ must be light \cite{unificaxion} (as discussed in Sec. \ref{sec:minimal-models}, however, we exclude models with $V$ altogether).

On the right of Fig. \ref{fig:gut} we show models as continuous half lines stemming from the origin, whose slopes and directions depend on the model content as described just above. 
The models shown are representatives of all the models among the viable ones 
of Tab. \ref{tab:all-models} 
whose lines cross the green region, which means that they may improve the unification according to our criterion for some $M_H$. Replacing one or more of the Dqs with their tilded versions one obtains models with the same lines, that may or may not be viable depending on dimension five operators or the existence of a viable DMC. In total, we count twentyfour models that pass this selection.
The value of the GUT coupling must be checked in all cases: typically, requiring that $\alpha_{\text{GUT}}$ is perturbative poses a lower limit on the value of $M_H$
, as can be seen the right panel of Fig. \ref{fig:gut_examples}.
As $M_H$ becomes larger, points along the lines are scanned farther and farther away from the origin, and eventually they leave the green area. This poses an upper limit on the heavy fermion scale $M_H$, depending both on the Dq content and on $\NDC$.

To give a concrete example, let us consider the model $Q\oplus E \oplus \widetilde{D} \oplus \widetilde{L}$ with $\NDC=3$. This model would be viable even without the fundamental scalar, as Yukawa couplings with the Higgs field suffice in breaking the species symmetries. However, it does not produce a successful exact unification, and indeed neither is it reported in \cite{CDM} nor in our Tab. \ref{tbl:GUT}.\footnote{Actually, in \cite{CDM} there is no mention of this model because it has $N_{\text{DF}}=12$ and no model is considered with a number of DFs larger than ten in that work. One can easily be convinced that it would indeed be a \emph{golden} model in their language as all accidental symmetries except for DB number are explicitly broken at renormalizable level.} Therefore it is a genuine case in which relaxing the criterion for unification makes the model successful. {Taking 
$M_H = \SI{e12}{\giga\electronvolt}$, we find
\be
 \MGUT =  \SI{7.4e14}{\giga\electronvolt} \quad \text{and} \quad \alpha_{\text{GUT}} = 5.9\times 10^{-2} \sim \frac{1}{17}\,.
\ee
The running of the SM couplings in this model are shown in the left panel of Fig. \ref{fig:gut_examples}.

\begin{figure}[!htt]
    \centering
    \includegraphics[width=0.45\textwidth]{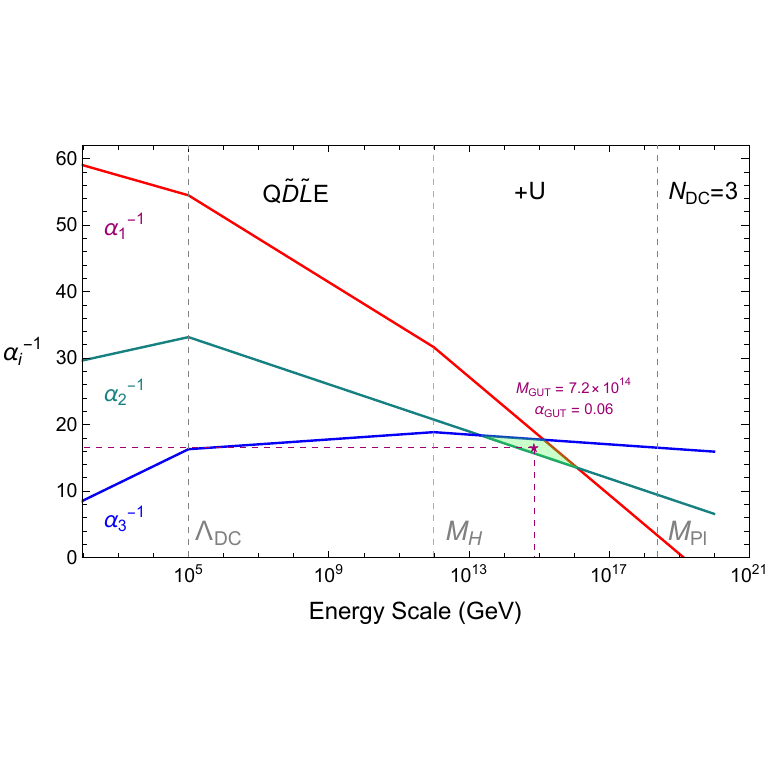}
    \hfill
    \includegraphics[width=0.45\textwidth]{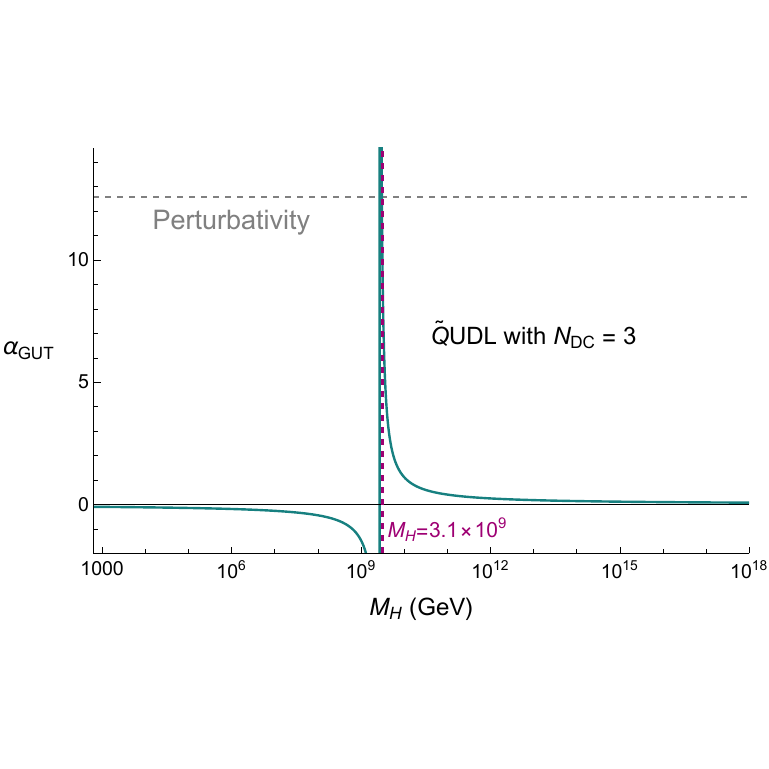}
      \caption{{\bf Left}: Running of the SM gauge couplings in the model $Q \oplus \widetilde{D} \oplus E \oplus \widetilde{L}$ with $\NDC=3$ 
      and $M_H=\SI{e12}{\giga\electronvolt}$. The purple star inside the highlighted triangle is its barycenter, whose coordinates correspond to the GUT scale and the inverse of the GUT coupling. {\bf Right}: Value of the GUT coupling as a function of $M_H$ in the model $\widetilde{Q} \oplus U \oplus D \oplus L$ with $\NDC=3$
      . The coupling is negative at first, and then non-perturbative until the value ${M_H} \sim 3.1 \times 10^9$ is reached, so the model cannot be considered successful for smaller values of the heavy Dq mass scale.
    }
    \label{fig:gut_examples}
\end{figure}

%% file: pheno/pheno.tex
Composite DM setups present very rich phenomenologies. As far as direct detection is concerned, the DM may have sizeable electromagnetic dipole moments because of its charged consitutents \cite{CDM}. Furthermore, they naturally feature light particles (dark pions and dark glueballs) interacting in number changing interactions, which produce non-standard cosmologies \cite{cannibal, phasesofcannibal, gluequark}, and may be testable in indirect detection experiments \cite{Dondi:2020}. Being some of these light states potentially charged under the SM, they may be observable at colliders \cite{darkmesonsLHC,ATLAS:darkmesons}. For general features of the phenomenology of models of accidental DM we refer the reader to previous literature (for the weakly coupled scenario, see \cite{WCD}). Here, we shall focus instead on the impact that the dark scalar $\phi$ has on phenomenology, both directly, by looking at dark states that contain $\phi$ as a constituent, or indirectly through the interactions it mediates, which break the accidental symmetries of the dark sector in a special way, transferring the SM accidental symmetries to the Dqs as envisioned in Tab. \ref{tab:accidental-numbers-recap}.
\subsection{Direct Detection of Hybrid Dark Matter Candidates}\label{sec:dd}
\input{pheno/dd}

\subsection{Dark Pions at Colliders}
\input{pheno/colliders}

\subsection{Dark Meson Leptoquarks}
\input{pheno/leptoquarks}
\subsection{Lepton Flavor and CP Violation}
\input{pheno/lepton-flavor-violation}

%% file: pheno/dd.tex
As pointed out in \cite{CDM}, the typical DM-nucleon cross sections for weak interactions \cite{MDM} are too weak to be detected at current DD experiments (see also \cite{lastWIMP}), and the best hope for DD would be interactions with photons through the electromagnetic dipoles of DBs. Such interactions only arise at the level of dimension six for scalar DM and of dimension five for Dirac DM, but spin one DM may have such interactions at the renormalizable level. Another possibility is to directly observe Higgs-mediated interactions between the nucleons and the DM, which naturally interact with the Higgs boson as its constituents have the tree-level Yukawa interactions of Tab. \ref{tab:yukawa}.

An interesting possibility for the detectability of our setup is to exploit the scalar portal $\lambda_{\phi\,H} \left|\phi\right|^2 \left|H\right|^2 \supset \lambda_{\phi\,H}\,v \left|\phi\right|^2 h$. Even in models in which no Yukawa couplings with the Higgs is allowed, if the DM is a hybrid state containing both fermionic and scalar Dqs (see Sec. \ref{sec:hybrid} for a discussion) it interacts with nuclei through this portal with the exchange of one Higgs boson. The DM also inherits a quartic coupling with two Higgs boson legs, which however mediates interactions with nuclei that are suppressed both by a loop factor and by the small coupling oh the Higgs with the nuclei. For Dirac DM, unless its gyromagnetic factor is especially small or $\lambda_{\phi H}$ is especially large, this interaction will be subdominant with respect to the dipole interaction. For scalar DM, however, the dipole interactions are suppressed, and this interaction may be more important.

Considering the weakly coupled scenario in this case, with $M_\phi \simeq M_\Psi \gg \LDC$, the (dimensionless) coupling of a hybrid Dirac DM to the Higgs boson can be computed by matching the matrix elements of the energy-momentum tensor on the hybrid DM states: 
\begin{equation}
    \lambda_{h\text{\tiny{-DM}}} = v \frac{\partial M_{\text{DM}}}{\partial\left(H^\dagger H \right)} = \frac{\lambda_{\phi H} v}{2 M_{\text{DM}}} \left\langle DM \right| \phi^\dagger \phi \left | DM\right\rangle =  \lambda_{\phi H} N_\phi \NDC \frac{v}{M_{\text{DM}}}\,.
\end{equation}
For a scalar hybrid DM, the dimension one coupling is obtained by multiplying the above equation by $2 M_{\text{DM}}$. The spin-independent (SI) cross-section for DD is then, for DM with any spin,
\be\label{eq:higgs-mediated-xsec}
\sigma_{\text{SI}} = \frac{\lambda^2_{h\text{\tiny{-DM}}} m_{\mathcal{N}}^4 f^2_{\mathcal{N}}}{2 \pi m_h^4 v^2} = \frac{\lambda^2_{\phi H} N_\phi^2 \NDC^2 m_{\mathcal{N}}^4 f^2_{\mathcal{N}}}{2 \pi m_h^4 M_{\text{DM}}^2}\,,
\ee
where $ f_{\mathcal{N}}\sim0.3$ is a nuclear form factor, $m_{\mathcal{N}}$, $m_h$, and $M_{\text{DM}}$ are the masses of the nucleon, Higgs boson, and DM, respectively. The LUX-ZEPLIN bound \cite{LZ} on the SI DD cross-section translates into $\lambda_{\phi H} < 30 \frac{4}{\NDC} \frac{1}{N_\phi} \left(\frac{M_{\text{DM}}}{\SI{100}{\tera\electronvolt}}\right)^\frac{3}{2}$, showing the elusiveness of this scenario.


%% file: pheno/colliders.tex
\label{sec:colliders}

 \begin{figure}[!t]
\centering
\begin{tikzpicture}
    \begin{feynman}
        \vertex (i1) {\(\)};
        \vertex[right=of i1] (m1);
        \vertex[right=of m1] (f1) {SM};
        \vertex[below = of i1] (i2) {\(\)};
        \vertex[right=of i2] (m2);
        \vertex[right=of m2] (f2) {SM};

        \diagram*{
        {[edges=fermion]
            (m1) -- (f1),
            (f2) -- (m2),
        },
        (i1) -- [fermion, very thick] (m1),
        (m2) -- [fermion, very thick] (i2),
        (m1) -- [scalar,very thick, edge label=\(\phi\)] (m2),
        };

        \draw[decoration = {brace},decorate] (i2.south west) -- (i1.north west) node [pos=0.5,left] {\(\pi_D\,\)};
    \end{feynman}
\end{tikzpicture}
\caption{Interaction between a dark pion and two SM fermions. Thick continous (dashed) lines represent fermionic (scalar) Dqs.}\label{fig:pi-prod}
\end{figure}
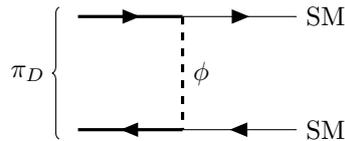
 In the strongly coupled scenario the DBs, dark vector mesons, and dark glueballs are much too heavy to be produced at a collider; however, D$\pi$s may be much lighter than the DM, and thus are the most promising to be probed. According to Eq. \ref{eq:pion-mass}, the gauge contribution gives roughly $ m_{\pi_D}\sim 0.1 \,\LDC \sim  \,\SI{10}{\tera\electronvolt}$ for charged D$\pi$s. If we assume that the Dq masses are some factor of a hundred smaller than the confinement scale $M_\Psi \sim \frac{\LDC}{100} \sim \SI{1}{\tera\electronvolt}$ (analogous to the first generation of QCD quarks), the contribution from the constituent masses is of the same order. We take this to be the benchmark value for the D$\pi$ masses.

Dark mesons without species number have anomalous couplings to vector bosons
, allowing to produce them through vector boson fusion. Any dark meson may also be pair produced through their EW interactions or through the mixing of the dark $\rho$ with the EW bosons \cite{darkmesonsLHC}. The novelty of our setup is that thanks to the new Yukawa interactions mediated by the scalar Dq $\phi$ the dark pions can decay directly to SM fermions through the interaction of Fig. \ref{fig:pi-prod}.

Because of the chiral structure of the SM, Dqs couple to SM fermions with specific chiralities, and depending on the constituents of the dark meson, the decay rates may or may not be chiral suppressed. For instance in a model containing $L \oplus E$, the dark mesons transforming as $\left(\mathbf{1},\mathbf{2}\right)_{-\frac{3}{2}}$ couple to $\ell^i_L + e^{i\,-}_R$, so the decay rate features the helicity suppression by a small SM lepton mass, in the same fashion as the QCD pion decay. Indeed in both cases the decay proceeds through an axial vector current. The same is true for dark pions without species number in any model. It is not the case for instance in models containing $L \oplus \widetilde{E}$, in which the dark mesons transforming as $\left(\mathbf{1},\mathbf{2}\right)_{-\frac{1}{2}}$ couple directly to $\overline{\ell}^{\,i}_R + e^{i,-}_R$ and decay without chiral suppression, through the dark psuedoscalar density $J_D^5$. We can estimate the rate as
\be
\Gamma\left(\pi_D \to \text{SM}\,\text{SM}\right) = \left(G^{L\,\widetilde{E}}_D\right)^2 \frac{m_\pi \LDC^4}{128 \pi^2}\,, 
\ee
where we defined the strength of the four fermion interaction to be
\be
G^{L\,\widetilde{E}}_D = \frac{y_L y^\ast_{\widetilde{E}}}{\LDC^2}\,.
\ee
 As this decay width is not chiral-suppressed it may be rather large, and portions of the parameter space of some models in which the D$\pi$s are especially light are likely to be excluded already by existing searches. On the other hand, future colliders may produce a large number of these states resonantly. For instance, under the reasonable assumptions on the mass scales described above, a muon collider with a realistic beam energy spread \cite{mucoll} would produce these states with cross sections at the level of a few to tens of $\si{\femto\barn}$ in the Breit-Wigner approximation. 

In most models, as all the spurious species symmetries are broken at renormalizable level, the decay of the mesons is prompt on collider scales. If, on the contrary, displaced vertices are observed, one can deduce that the D$\pi$s arise from a model with residual species symmetries broken only at higher dimension level. The only models among those we found that have this characteristic are those that feature $N$ but not $L$ in the light Dq spectrum, for which the breaking of the species number happens through the leftmost operator in Eq. \eqref{eq:Nbreak3colors}.

%% file: pheno/leptoquarks.tex
Models where both Dqs with baryon number and Dqs with lepton number are present in the light Dq spectrum produce dark mesons with both baryon and lepton number. They are leptoquarks (LQ) candidates, as they will couple to both SM leptons and quarks through the dark scalar.

In the strongly coupled scenario ($\Lambda_{DC}\sim \SI{100}{\tera\electronvolt}$), 
the D$\pi$s have masses around $\sim \SI{10}{\tera\electronvolt}$ (see above), out of current reach of dedicated LHC searches \cite{cms-lq}. Let us take, as examples, models that include $Q\oplus \widetilde{L}$ in the light Dq spectrum. In these cases two LQs appear, with quantum numbers $(\overline{\three},\one)_{\frac{1}{3}}\oplus (\overline{\three},\three)_{\frac{1}{3}}$, 
usually referred to in the literature as $S_1$ and $S_3$. 
In this scenario the quadratic splitting between $S_{1}$ and $S_{3}$ dark meson masses is determined only by gauge interaction since the constituent Dq are the same, and is of order $ m_{S_3}^2-m_{S_1}^2\sim \left(\SI{5}{\tera\electronvolt}\right)^2$. 
The $S_{3}$ is an example of non-genuine LQ (for a reference on leptoquarks see \cite{LPref}), that is, a LQ that forms baryon-number breaking terms with SM fermions:
\begin{equation}
    \mathcal{L} \supset y_{ij}q_{i}(\epsilon\,\vec \tau \cdot \vec S_{3})l_{j} + z_{ij} q_{i}(\epsilon\,\vec \tau \cdot \vec S_{3})^{\dagger}q_{j} + h.c.
    \label{eq:LLQ}
\end{equation}
It is important to notice that the baryon number breaking term is suppressed as $\sim \frac{1}{M_{GUT}^2}$, because in the UV theory where Dqs and the scalar are the fundamental degrees of freedom, the baryon symmetry is preserved, so the only contribution comes from the GUT scale. On the other hand, the couplings $y_{ij}$ are only suppressed by the mass of the dark scalar, or the confinement scale. 



The phenomenology of the dark meson LQs in composite DM theories should be subject of further investigations. For instance, the mixing between the charge $Q=\frac{1}{3}$ states in the two LQs mentioned above, which proceeds through their couplings with the Higgs boson, depending in turn onnly on the Yukawa couplings of the model, may be exploited to ameliorate the muon $g-2$ problem \cite{LQS1S3}. 

%% file: pheno/lepton-flavor-violation.tex
\label{sec:lfv}

As described in Sec. \ref{sec:outline}, the presence of the scalar Dq allows to write 
the Yukawa couplings of Tab. \ref{tab:yukawa}, which may violate flavor and $CP$. Here we recast the analysis of \cite{bearable-compositeness} and discuss the consequences on the lepton sector. The relevant Dqs to consider are $L$ and $E$ (or $\widetilde{E}$), that couple to the SM particles via the Yukawa terms $L^c\phi l$ and $E^c\phi e^c$ ($\widetilde{E} \phi^\dagger e^c$). We write the Yukawa coupling joining a SM fermion $\psi^{\text{SM}}$ 
 and a fermionic Dq $\Psi$ through the dark scalar as $y^{\psi^{\text{SM}}}_\Psi$. If we are describing, for example, interactions involving left-handed muons or muonic neutrinos
 , the relevant coupling is $y^{\ell_2}_L$. In the case of $L\oplus\widetilde{E}$, there is also the coupling $y_{L\widetilde{E}} L^c H \widetilde{E}$.
 
We estimate the contributions of our dark sectors to the Wilson coefficients of operators in the SM EFT through spurionic arguments and na\"ive dimensional analysis. We consider the strongly coupled scenario and assume for simplicity that all resonances of the dark sector are controlled by just one massive parameter, which we take conservatively to be the smallest mass scale of the theories, namely the D$\pi$ mass scale 
$m^* \sim m_{\pi_D} \sim \SI{10}{\tera\electronvolt}$.
We thus neglect the mass differences between the various dark hadrons. We assume that all the dark Yukawa couplings are of the same order.
Furthermore, we make the same assumptions as in \cite{bearable-compositeness}, namely that the UV theory contains the full basis of dimension six operators at the scale $\mu=m^\ast$, and that renormalization group effects and the interference between the operators can be ignored. 

We use the experimental bounds \cite{bearable-compositeness} on processes that violate flavor (such as lepton decays $\mu\to e e e$) as well as on observables such as the electic dipole moment (EDM) to derive upper bounds on the Yukawa couplings. The latter allow to derive constraints on both the real and the imaginary parts of the products of Yukawa couplings, as they violate CP. In our models these observables depend on the product of several of the Yukawa couplings, and in general it is not possible to put bounds on any one coupling. We thus assume that they are all of the same order in order to estimate the bounds. 

The relevant dimension six operators (written as in the Warsaw basis \cite{Grzadkowski_2010}) are enlisted in Tab. \ref{tab:dim6op}. We take a flavor-diagonal basis for our fields

\begin{table}[ht!]
    \centering
    \caption{Relevant dimension-six operators for the flavour violating processes. Lower-case latin indices are lepton family indices.}
    \vspace{2mm}
    \scalebox{1.2}{
    \begin{tabular}{c|c}
        Effective operator &  Wilson coefficient\\
        \hline
        \hline
        $Q^{ij}_{e \gamma} = (l_i \sigma^{\mu\nu} e^c_j) H^\dagger F_{\mu\nu}$ &  \makecell*{$\frac{C^{e\gamma}}{\Lambda^2} =\frac{c^{e\gamma}_{ij}}{16 \pi^2 m^{*2}}e y_{L}^{i*}y^{j}_{\widetilde{E}}y_{HL\widetilde{E}}^{*}$ \\ 
        or \\
        $\frac{C^{e\gamma}}{\Lambda^2} =\frac{c^{e\gamma}_{ij}}{256 \pi^4 m^{*2}}e Y_{\text{SM}}^{k\,k} y_{L}^{i\,\ast} y_L^{k} y^{k\,\ast}_{E} y_E^{j}$}\\
        \hline
        $Q^{ijmn}_{ll} = (l_i\sigma_\mu l^{\dagger}_j) (l_m\sigma^\mu l^{\dagger}_n)$ & $\frac{C^{ll}}{\Lambda^2} = \frac{c^{ll}_{ijmn}}{{16 \pi^2} m^{*2}}y^{i*}_L y^{j}_L y^{m*}_L y^{n}_L$ \\
        $Q^{ijmn}_{le} = (l_i \sigma_\mu l^{\dagger}_j) (e^c_m\sigma^\mu e^{c\dagger}_n)$ & $\frac{C^{le}}{\Lambda^2} = \frac{c^{le}_{ijmn}}{{16 \pi^2}  m^{*2}}y^{i*}_L y^{j}_L y^{m*}_E y^{n}_E$ \\
        $Q^{ijmn}_{ee} = (e^{c}_i\sigma_\mu e^{c \dagger}_j) (e^{c}_m\sigma^\mu e^{c\dagger}_n)$ & $\frac{C^{ee}}{\Lambda^2} = \frac{c^{ee}_{ijmn}}{{16 \pi^2}  m^{*2}}y^{i*}_E y^{j}_E y^{m*}_E y^{n}_E$ \\
    \end{tabular}}
    \label{tab:dim6op}
\end{table}

Processes arising from four-fermions operators, such as $\mu \to e e e$,
do not allow to put any stringent bound as all coefficients satisfy the constraints with $y_D$ as large as $\sim \mathcal{O}(1)$. The same is true for almost all the processes arising from two-fermion operators, which include the decay of muons and tau leptons into lighter leptons and photons, as well as the EDMs. The only exception are the coefficients of the operator $Q^{11}_{e \gamma}$, which is responsible for the EDM of the electron 
\begin{equation}
    \frac{C^{e\,\gamma}_{11}}{\Lambda_{UV}^2}(l_i \sigma^{\mu\nu} e^c_j) H^\dagger F_{\mu\nu}
\end{equation}
This bound on a $CP$ violating process is the most stringent: at $m^\ast = \SI{10}{\tera\electronvolt}$ it gives $\text{Im} C^{e\,\gamma}_{11} < \SI{3.8 e-10}{}$ \cite{bearable-compositeness}.

In the absence of a symmetry suppressing the violation of CP,\footnote{Since $C^{e\,\gamma}_{ij}$ depends on all six $y^i_{L}$,$y^i_{E}$, in order to have enough CP suppression one needs a symmetry breaking pattern allowing for the redefinition of six phases. The pattern in general will be 
\be
U(1)^3_l \otimes U(1)^{n_{\text{sp}}+1}_D \to U(1)_l \otimes U(1)_{\text{DB}}\,,
\ee
where $n_{\text{sp}}$ is the number of fermionic Dq species, and the plus $1$ is the species number of the scalar $\phi$. The total number of unphysical phases is then $n_{\text{sp}}+2$, from which we find that models with four or more fermionic species are free from CP violation in the lepton sector at the two loop level.} one finds, in the case of $L\oplus E$, in which the operator is generated at two loops, that the bound translates into $\text{Im}\left[(y_L)^2 (y_{\widetilde{E}})^2\right] \lesssim \SI{4.5e-3}{}$, which is passed with $y_{L,\,E} \lesssim \SI{0.26}{}$. In the case of $L\oplus \widetilde{E}$, instead, the operator is generated at one loop, and the bound is more stringent: $y_L  y_{\widetilde{E}} y_{L\,\widetilde{E}} \lesssim \SI{2e-7}{}$, which requires $y_L, y_{\widetilde{E}}, y_{L\,\widetilde{E}}  \lesssim \SI{6e-3}{}$.
Even if CP violation is suppressed, the first operator in Tab. \ref{tab:dim6op} poses the most stringent bound, from \cite{MEGII}
\be
\text{BR}\left(\mu \to e \gamma\right) < \SI{3.1e-13}{} \quad \text{at 90\% CL}\,,
\ee
translating into $y_L, y_E \lesssim 0.7$ and $y_L, y_{\widetilde{E}}, y_{L\,\widetilde{E}} < 0.02 $ for $L\oplus E$ and $L\oplus\widetilde{E}$, respectively.

The only question then becomes if these upper bounds on the Yukawa couplings are in conflict with the requirement that the D$\pi$s decay before BBN. In models featuring $L\oplus\widetilde{E}$, as discussed in Sec. \ref{sec:colliders}, the decays mediated by the scalar Dq are not chirally suppressed, and are thus quite fast. With the usual assumptions of $\LDC \sim \SI{100}{\tera\electronvolt}$, $m_\pi \sim \SI{10}{\tera\electronvolt}$, $M_\phi \gtrsim \LDC$, the D$\pi$s decay before BBN with yukawa couplings as small as $\sim 10^{-6}$. In models with $L\oplus E$, on the other hand, the decay is chiral suppressed and lifetimes may be larger. Indeed larger yukawa couplings of order $10^{-5}$ are required. In neither case does this simplified analysis lead to an evident exclusion. We leave a more detailed analysis on selected models to a future work.

%% file: conclusions.tex
In this work we 
studied extension of the SM with \emph{dark color} sectors with special unitary gauge groups. The novelty with respect to previous literature on the subject is that we included both fermionic and scalar Dqs.
The advantage of such dark sectors in general is that they constitute fundamental, UV complete theories in which all the desired properties of the DM -- most importantly its stability on cosmological time-scales and its lack of color and hypercharge interactions -- are a consequence of merely the quantum numbers of the fundamental degrees of freedom and the ensuing accidental symmetries of the theory. Indeed they naturally provide a plethora of states, the analog of the QCD hadrons, the lightest of which are stable as a consequence of the accidental symmetries of the theory. In this work we considered the lightest DBs as the DMC. Since the SM gauge interactions lift the masses of the charged DB it comes automatically that the DM is uncharged if the quantum numbers of the constituents allow for it. Yet one has a number of charged partners, some of which may be light; these states may be exploited to test the models in various experiments.

It is the ease with which one can break undesired accidental symmetries, as needed to avoid spurious stable states whose existence would conflict with observations, that drove us to include the scalars. Indeed we found many more models to be viable with respect to the previous literature. We postponed entirely the discussion of the hierarchy problem that one comes across when including fundamental scalars in the theory.
The standard configuration we assumed is one in which the fermionic Dqs are lighter than the confinement scale, analogously to the lightest quarks in QCD, realizing a strongly coupled scenario. Cosmology in this case fixes the mass of the DM to be around a few hundreds of $\si{\tera\electronvolt}$. 
 We also discussed, however, the possibility of shuffling the order of the masses to realize weakly-coupled scenarios.

We analyzed the models in the context of an $SU(5)$ grand unification scheme. We followed a recent proposal by \cite{verma} and considered a \emph{relaxed} criterion for unification: we consider as successfully unifying models in which the unification of the SM couplings is \emph{better} than in the SM, yet possibly still imperfect. Practically speaking, rather than having the three couplings to match precisely at a certain energy scale, we accept models in which the area of the triangle drawn by the running in the energy scale-inverse coupling plane is smaller than in the SM. Overall, we find twenty-four models to be successfully unifying, to be compared with essentially a single model in previous literature. This is both thanks to a larger set of models as allowed by the existence of the scalar and thanks to this relaxed criterion.  

In the section dedicated to phenomenology, we discussed aspects of the testability of the models. Given that many features are common with those discussed in the existing literature on accidental composite dark matter, we focused exclusively on how phenomenology is impacted by the presence of the scalar, with special attention to direct detection experiments and particle colliders. The most interesting features are provided by Yukawa couplings between the Dqs and the SM fermions mediated by the dark scalar. As far as collider phenomenology is concerned, they allow for the production and subsequent decay of dark pions directly to SM fermions.
Another crucial consequence of these Yukawa interactions is the transfer of the SM accidental symmetries to the dark sector, 
with the result that some of the dark states possess both SM lepton and baryon numbers, making them LQs candidates: the extensive literature on LQs suggests several phenomenological application to be studied on specific models.

A feature of any of the models here discussed is the existence of \emph{hybrid} dark states containing both fermions and scalars. It is beyond the scope of this work to determine the dynamics of the formation of these states and to assess the hierarchy between them and the "regular" dark states, which is challenging especially in strongly coupled scenarios. We have however argued that it is possible under some conditions that even the DMC may be of hybrid nature. This would lead to Higgs-portal interactions with nucleons to be constrained at DD experiments. Interestingly if these interactions are leading with respect to dipole/charge-radius interactions, hybrid DM is quite elusive of stringent DD bounds.

Hybrid sates with two Dqs (hybrid dark mesons) are fermionic in nature and mix with the SM fermions. The new Yukawa couplings that govern this mixing violate the SM symmetries in general. We discussed the case of models with Dq that carry lepton number, in which LFV and CP violating processes are mediated by these interactions, to translate the stringent bounds on the lepton sector into bounds on the parameters of the models. There are a large number of these parameters, which always enter in the observables in products of three or more, hence it is impossible at the time to constrain single parameters. However, under certain reasonable assumptions, it is possible to establish whether some models are already excluded by observations. A simplified analysis shows that this is not the case.
It would be compelling to delve deeper and explore these phenomenological possibilities in future works.

%% file: model-building/examples.tex
In this appendix we show the details of one model that passes the selection, one model whose viability is spoiled by dimension five operators peculiar to our setup
, and one model that does not pass the selection. In all cases we consider only two light Dq species and make use of the Mathematica package LieART \cite{Mathematica,LieART} to compute the decomposition of the DF representation into SM representation. The procedure can be iterated to decompose the DF representation of models with more than two light Dq species. 

\subsection*{a)  \texorpdfstring{$Q \oplus \widetilde{D}$}{Q+Dtilde}}
\label{par:QDtilde}

This model has two species and $N_{DF} = 9$. Species numbers are broken thanks to the dark Yukawa couplings $y_Q Q^c \phi q$, $y_{\widetilde{D}} \widetilde{D} \phi^\dagger d^c$, and $y_{Q\widetilde{D}} Q^c H \widetilde{D}$. The Landau poles constraint gives $\NDC\le4$. Let us consider $\NDC=3$. The DB multiplet to decompose is the representation \textbf{240} of $SU(9)_{DF}$, under $SU(3)_c \otimes SU(2)_L \otimes U(1)_Y$:

\begin{equation} \label{eq:QDt}
    \begin{split}
        \textbf{240} = & (\one,\one)_0 \oplus (\one,\two)_{\pm 1/2} \oplus (\one,\three)_0 \oplus (\textbf{8},\one)_{2\times 0, -1} \oplus (\textbf{8},\two)_{2\times -1/2, 1/2} \\& 2(\textbf{8},\three)_0 \oplus (\textbf{8},\four)_{1/2} \oplus (\textbf{10},\one)_0 \oplus (\textbf{10},\two)_{\pm 1/2} \oplus (\textbf{10},\three)_0
    \end{split}
\end{equation}
The fermionic DMC is the singlet or the neutral state of the triplet in $QQ\widetilde{D}$:
\begin{equation}
    QQ\widetilde{D} \sim (\one,\one)_0 \oplus (\one,\three)_0 
\end{equation}
Another valid candidate to include in the classification is the hybrid $QQ\widetilde{D}\phi$, obtained by attaching a scalar to the fermionic one and raising $\NDC$ to $4$.

\subsection*{b)  \texorpdfstring{$E \oplus L$}{E+L}}

This model has $N_{DF} = 3$ and it avoids the Landau poles for $N_{DC} = 3,4,5$.
All species numbers are broken by Yukawa terms $E^c\phi e^c + h.c$ and $L^c\phi l + h.c$, and the D$\pi$s are unstable.
For $N_{DC}=3$, the DB multiplet to decompose under $SU(2)_L\otimes U(1)_Y$ is:   
\begin{equation}
        \textbf{8} = \one_0 \oplus \two_{\pm 3/2} \oplus \three_0
\end{equation}
The possible DMCs are 
\begin{equation}
    ELL \sim \one_0 \oplus \three_0
\end{equation}
However, the dimension five operator 
\begin{equation}
     LH^{\dagger}E\phi 
\end{equation}
breaks the DB number explicitly for $\NDC=3$. 
Regarding hybrid DM candidates, given the bounds from the Landau poles, one finds the singlets $ELL\phi$ ($N_{DC} = 4$) and $ELL\phi^2$ ($N_{DC} = 5$): in these cases there is no dimension five operator that breaks the DB number.

\subsection*{c) \texorpdfstring{$D \oplus L$}{D+L}}

In this model we can build no Yukawa coupling with the Higgs, but the terms 
$L^c\phi l + h.c$ and $D^c\phi d^c + h.c$ 
 break respectively $U(1)_L$ and $U(1)_D$ species numbers, as well as the $\phi$ species number. The Landau poles constraint allows to build models up to $N_{DC} = 9$, and a state with $Y=0$ exists with $N_{DC} = 5$, schematically 
$DDDLL$, which would be stable thanks to $U(1)_{DB}$. Nonetheless, this state does not belong to 
the lightest DB multiplet.
In fact, for $N_{DC} = 5$ the Young tableau for DF and spin is 
\begin{equation}
    \ydiagram{3,2}\,,
\end{equation}
which is the representation $\mathbf{175'}$ of $SU(5)_{\text{DF}}$, whose decomposition under $SU(3)_c \otimes SU(2)_L \otimes U(1)_Y$ is
\begin{equation}
\begin{split}
    \mathbf{175'} = & (\one,\two)_{-5/2} \oplus (\overline{\three},\one)_{-5/3} \oplus (\mathbf{3},\mathbf{2})_{-5/6} \oplus (\overline{\three},\three)_{-5/3} \oplus \\ & (\mathbf{6},\mathbf{2})_{5/6} \oplus (\mathbf{\overline{6}},\mathbf{2})_{-5/6}  \oplus (\mathbf{8},\one)_0 \oplus (\mathbf{\overline{6}},\four)_{-5/6} \oplus (\mathbf{8},\three)_0 \oplus \\ & (\mathbf{\overline{10}},\three)_0 \oplus (\mathbf{15},\one)_{5/3} \oplus (\mathbf{\overline{15}},\two)_{5/6}\,.
\end{split}
\end{equation}
Since there is no state in the multiplet that is uncoloured and with $Y=0$, there is no viable DMC.\\





%% file: model-building/Minimal_extension.tex
We can extend models starting from the observation of the following property of the decompositions. Let's take a model of the form:
\begin{equation}
    \Psi = \Psi_{1} \oplus \Psi_{2}
\end{equation}
following the method exposed in \cite{CDM} the first DF decomposition is:
\begin{equation}
    SU(N_{DF})_{\Psi} \longrightarrow SU(N_{1})_{\Psi_{1}} \otimes SU(N_{2})_{\Psi_{2}} \otimes U(1)_{X}
\end{equation}
where $SU(N_{DF})_{\Psi}$ is the DF group of the whole model, $SU(N_q)_{\Psi_{1}}$ is the DF group of the species $\Psi_{1}$, $SU(N_2)_{\Psi_{2}}$ is the DF group of the species $\Psi_{2}$, and $U(1)_{X}$ is related to the hypercharge. 
We can represent this decomposition by means of Young tableaux:
\begin{equation}\label{eq:decEM}
\scalemath{0.72}{\Yvcentermath1\yng(2,1)= \left(\yng(1,1),\yng(1)\right)_{2Y_{1}+Y_{2}} \oplus \left(\yng(1),\yng(1,1)\right)_{2Y_{2}+Y_{1}}\oplus \left(\yng(2),\yng(1)\right)_{2Y_{1}+Y_{2}}\oplus \left(\yng(1),\yng(2)\right)_{2Y_{2}+Y_{1}}\oplus\left(\yng(2,1),1\right)_{3Y_{1}}\oplus\left(1,\yng(2,1)\right)_{3Y_{2}}}
\end{equation}
in the last two terms the same tableau appears that corresponds to the $SU(N_{DC})_{\Psi}$ representation, but this time for the $\Psi_{1}$ and $\Psi_{2}$ representations. This kind of terms always appear in the decompositions. This procedure can be iterated if we have more than 2 light Dq species: for example with three Dq representation
\begin{equation}
    \Psi = \Psi_{1}\oplus \Psi_{2} \oplus \Psi_{3}
\end{equation}
we can group Dqs as $\Psi = \Psi_{M}\oplus \Psi_{3}$, where $\Psi_{M} = \Psi_{1} \oplus \Psi_{2}$, and  make the decomposition exposed above,  then repeat the procedure for $\Psi_{M}$. Let's focus on models with an arbitrary number of Dq species of the form:

\begin{equation}
    \Psi= \Psi_{M}\oplus\Psi_{S}
\end{equation}
where $\Psi_{M}$  is a minimal model and $\Psi_{S}$ is a spectator Dq, that is, a Dq that is not a constituent of the DMC of the model. We say that $\Psi$ is an extension of the minimal model $\Psi_M$. The last two terms in Eq. \eqref{eq:decEM} imply that we only need to study  minimal models: if we add a spectator light Dq $\Psi_{S}$ (or a set of spectators Dq), we use the first step in order to decompose the total DF group into the product of the DF groups of the minimal model and the spectator Dqs:
\begin{equation}
    SU(N)_{DF} \longrightarrow SU(d_M)_{M} \otimes SU(d_S)_{S} \otimes U(1)_{Y}
\end{equation}
where $d_M$ ($d_S$) is the dimension of the SM representation of $\Psi_{M}$ ($\Psi_{S}$).  In the decomposition of representations of $SU(d_M)_{M} \otimes SU(d_S)_{S}$ we always get the right representation for $SU(d_M)_{M}$ in order to get the DMC associated to the minimal model that respect the Fermi statistic as we saw in the example with the Young tableau.

It is also possible that the extended model is the extension of two different minimal model at the same time, in this case both candidates associated to the two different minimal models can be realized. Let us consider two minimal models, e.g. $Q \oplus \widetilde{D}$ and $\widetilde{D} \oplus \widetilde{U}$. What can we say about $Q\oplus \widetilde{D} \oplus \widetilde{U}$?
Let us consider the two models separatly:
\begin{itemize}
    \item $Q\oplus \widetilde{D}$:\\
    The DF decomposition for $N_{DC}=3$ made in the last section gives the result in Eq. \eqref{eq:QDt}, with DMC in  $QQ\widetilde{D}\sim(\one,\one)_{0}\oplus(\one,\three)_{0}$. We will refer to the full SM decomposition of the dark flavor representation (the right hand side of Eq. \eqref{eq:QDt}) so obtained as $M_{Q\oplus \tilde{D}}$.

\item $\widetilde{D}\oplus \widetilde{U}$:\\
There is a candidate for $N_{DC} = 3$. The first decomposition is $SU(6)_{\widetilde{D}\oplus \widetilde{U}}\longrightarrow SU(3)_{\widetilde{D}
}\otimes SU(3)_{\widetilde{U}}\otimes U(1)_{Y}$. The second decomposition is trivial since the DF of $\widetilde{D}$ comes all from the color representation, and the same holds for $\widetilde{U}$. So the decomposition is:

\begin{equation}
    \mathbf{70} = (\one,\one)_{0,1}\oplus (\mathbf{8},\one)_{2 \times 1, 2 \times 0, -1, 2} \oplus (\mathbf{10},\one)_{0,1} = M_{\widetilde{D}\oplus\widetilde{U}}
\end{equation}

the DMC is inside $\widetilde{D}\widetilde{D}\widetilde{U} = (\one,\one)_{0}$ multiplet.
\end{itemize}
Now in order to study the model $Q\oplus\widetilde{D}\oplus \widetilde{U}$ for $N_{DC} = 3$ we make the first decomposition in two different ways: first we decompose $SU(12)_{Q\oplus\widetilde{D}\oplus\widetilde{U}}\longrightarrow SU(9)_{Q\oplus\widetilde{D}}\otimes SU(3)_{\widetilde{U}}\otimes U(1)_{Y}$, so (omitting the hypercharge)
\begin{equation}
    \mathbf{572}= (\one,\mathbf{8})\oplus (\mathbf{9},\threebar)\oplus (\mathbf{9},\mathbf{6})\oplus (\mathbf{36},\three) \oplus (\mathbf{45},\three)\oplus (\mathbf{240},\one)
\end{equation}
The $(\mathbf{240},\one)$ is the one we decomposed in the minimal model $Q\oplus \widetilde{D}$. Second  we make the decomposition is $SU(12)_{Q\oplus\widetilde{D}\oplus\widetilde{U}}\longrightarrow SU(6)_{Q}\otimes SU(6)_{\widetilde{D}\oplus \widetilde{U}}\otimes U(1)_{Y}$. This must lead to the same result because it cannot depend on the way we factorize the decomposition. So the decomposition of the flavor representation in SM is:
\begin{equation}
    \mathbf{572}=(\mathbf{6},\mathbf{15})\oplus(\mathbf{15},\mathbf{6})\oplus(\mathbf{21},\mathbf{6})\oplus(\mathbf{6},\mathbf{21})\oplus(\mathbf{70},\one)\oplus(\one,\mathbf{70})
\end{equation}
the $(\one,\mathbf{70})$ is the representation we decomposed before in the $\widetilde{D}\oplus\widetilde{U}$ minimal model. This means that both candidates need to appear in the model $Q\oplus\widetilde{D}\oplus\widetilde{U}$. To check let us consider the decomposition of the DF representation for $Q\oplus\widetilde{D}\oplus\widetilde{U}$ with $N_{DC} = 3$ \cite{CDM}:
\begin{equation}
\begin{aligned}
    \mathbf{572}=M_{Q\oplus\widetilde{D}}\oplus (\one,\one)_{0,2\times 1}\oplus (\one,\two)_{2\times\frac{1}{2},\frac{3}{2}}\oplus(\one,\three)_{1}\oplus(\mathbf{8},\one)_{2\times0,4\times1,2}\oplus\\\oplus(\mathbf{10},\one)_{0,2\times1}\oplus (\mathbf{8},\two)_{4\times\frac{1}{2},2\times\frac{3}{2}}\oplus(\mathbf{10},\two)_{2\times\frac{1}{2},\frac{3}{2}}\oplus2\times(\mathbf{8},\three)_{1}\oplus(\mathbf{10},\three)_{1}\,.
    \end{aligned}
\end{equation}
Both $M_{Q\oplus \widetilde{D}}$ and $M_{\widetilde{D}\oplus \widetilde{U}}$ are contained in the multiplet. It would seem that one $(\mathbf{8},\one)_{-1}$ representation, contained in $M_{\widetilde{D}\oplus\widetilde{U}}$ is missing, but it is actually already included in $M_{Q\oplus \widetilde{D}}$. Indeed it is a $\widetilde{D}\widetilde{D}\widetilde{D}$ DB, which is common to both models.

Another way to extend minimal models is to add the Dq $N$ and increase $N_{DC}$ by one. With this procedure we can build a minimal model starting from another minimal model. For example starting from a model $Q\oplus \widetilde{D}$  for $N_{DC} = 3$ we can build a model  $Q\oplus \widetilde{D} \oplus N$ and $N_{DC} = 4$. The former model has $QQ\widetilde{D}$ as DMC, while the latter has $QQ\widetilde{D}N$. 
Of course one needs to make sure that this new model is not discarded because of dimension five operators that break the DB number or because it produces subplanckian Landau poles.
Note that this procedure cannot be iterated arbitrarily: for example iterating three times this procedure starting from the DF representation of the minimal model $Q\oplus \widetilde{D}$ we have in terms of Young tableaux:
\begin{equation} {\Yvcentermath1\yng(2,1)\longrightarrow \yng(2,2) \longrightarrow \yng(3,2) \longrightarrow \yng(3,3)}
\end{equation}
the last tableau with $N_{DC}=6$ is associated to a DF representation that, once we decompose it into SM representation, do not include any DMC of the form $QQ\widetilde{D}NNN$: this is because in the last tableau two anti-symmetrized  boxes have to correspond necessarily to two $N$ Dqs, which cannot, however, be antisymmetrized.

%% file: model-building/all-models.tex
\begin{table}[!p]
    \centering
    \renewcommand{\arraystretch}{1.8}
        \caption{Complete list of the viable models according to the criterion of Sec. \ref{sec:selection} with the exclusion of the models in which $N$ is light. It is worth stressing that this list is valid in the strongly coupled scenario. In the first column we show the $SU(5)$ representation to which the light Dqs (second column) belong. In the third column we show the allowed values for $\NDC$ and in the last column the potential DMCs for all the possibilities. Continues in Tab. \ref{tab:all-models-cont}}
    \label{tab:all-models}
    \vspace{2pt}
    \begin{tabular}{|c|c|c|c|}
    \hline
            $SU(5)$ &Model&$\NDC$&Dark Matter Candidate  \\
         \hline
       \rowcolor{yellow!25}
       &$D\oplus \widetilde{D}$&$4\le N_{DC}\le 7$& \makecell*{$D\widetilde{D} \phi^{\NDC-2}$, 
         $D\widetilde{D} D\widetilde{D}\phi^{\NDC-4}$, 
         $D\widetilde{D} D\widetilde{D} D\widetilde{D}\phi^{\NDC-6}$
         }\\
          \rowcolor{yellow!25}
          &$D\oplus \widetilde{D}\oplus L$&$4 \le N_{DC}\le 7$& \makecell*{
         $D\widetilde{D} \phi^{\NDC-2}$, 
         $D\widetilde{D} D\widetilde{D}\phi^{\NDC-4}$, 
         $D\widetilde{D} D\widetilde{D} D\widetilde{D}\phi^{\NDC-6}$
         }\\
          \rowcolor{yellow!25}
          &$L\oplus \widetilde{L}$&$4 \le N_{DC}\le 7$& \colorbox{yellow!25}{\makecell*{
         $L\widetilde{L} \phi^{\NDC-2}$, 
         $L\widetilde{L} L\widetilde{L}\phi^{\NDC-4}$, 
         $L\widetilde{L} L\widetilde{L} L\widetilde{L}\phi^{\NDC-6}$
         }}\\
          \rowcolor{yellow!25}
          &$L\oplus \widetilde{L}\oplus D$&$4 \le N_{DC}\le 6$& \colorbox{yellow!25}{\makecell*{
         $L\widetilde{L} \phi^{\NDC-2}$, 
         $L\widetilde{L} L\widetilde{L}\phi^{\NDC-4}$, 
         $L\widetilde{L} L\widetilde{L} L\widetilde{L}$
         }}\\
         \rowcolor{yellow!25}
         \multirow{-5}*{$\fivebar \oplus \five $}&$D\oplus\widetilde{D}\oplus L\oplus \widetilde{L}$&$ 4$&\colorbox{yellow!25}{\makecell*{$D\widetilde{D}\phi^2$,
        $L\widetilde{L}\phi^2$,
        $D\widetilde{D}D\widetilde{D}$, 
        $L\widetilde{L}L\widetilde{L}$, 
        $D\widetilde{D}L\widetilde{L}$}}\\
         \hline
         \rowcolor{orange!25}
         &$D\oplus U$&$ \le 4$&\colorbox{orange!25}{\makecell*{$DDU$, $DDU\phi $}}
         \\
          \rowcolor{orange!25}
         &$D\oplus U\oplus L$& $ 3$&$DDU$\\
         \rowcolor{orange!25}
        &$D\oplus U\oplus E$& $ 3$&$DDU$\\
        \rowcolor{orange!25}
        &$L\oplus E$&$ \le 5$&\colorbox{orange!25}{\makecell*{$LLE\phi$, $LLE\phi^2$}}\\
         \rowcolor{orange!25}
        \multirow{-5}*{$\fivebar \oplus \mathbf{10}$}&$L\oplus E \oplus D$&$ 4$&$LLE\phi $\\
        \hline
         \rowcolor{green!25}
         &$Q\oplus\widetilde{D}$&$\le 4$&\colorbox{green!25}{\makecell*{$QQ\widetilde{D}$,  $QQ\widetilde{D}\phi $
         }}\\
          \rowcolor{green!25}
        &$Q\oplus\widetilde{D}\oplus \widetilde{L}$&$ 3$&$QQ\widetilde{D}$\\
        \rowcolor{green!25}
        &$Q\oplus\widetilde{D}\oplus E$&$\le 4$&\colorbox{green!25}{\makecell*{$QQ\widetilde{D}$,  $QQ\widetilde{D}\phi $
        }}\\
        \rowcolor{green!25}
        &$Q\oplus\widetilde{D}\oplus \widetilde{L}\oplus E$&$ 3$&$QQ\widetilde{D}$\\  
        \rowcolor{green!25}
        \multirow{-5}*{$\five \oplus \mathbf{10}$}&$\widetilde{D}\oplus E\oplus U$&$ 3$&$\widetilde{D}EU$\\
        \hline
   \end{tabular}
\end{table}
\begin{table}[!hp]
    \centering
    \renewcommand{\arraystretch}{1.3}
    \caption{Continuation of Tab. \ref{tab:all-models}.}\label{tab:all-models-cont}
    \begin{tabular}{|c|c|c|c|}
    \hline
        $SU(5)$ &Model&$\NDC$&DMC  \\
         \hline
          \rowcolor{cyan!25}$\mathbf{10}\oplus\mathbf{\overline{10}}$&$E\oplus \widetilde{E}$&$4$&\colorbox{cyan!25}{\makecell*{$E\widetilde{E}E\widetilde{E}$, $E\widetilde{E}\phi^2$}}\\
        \hline
         \rowcolor{pink!25}
         &$D\oplus \widetilde{D}\oplus U$&$ 4$&\colorbox{pink!25}{\makecell*{$D\widetilde{D}D\widetilde{D}$,$D\widetilde{D}\phi^2$, $DDU\phi$}}\\
          \rowcolor{pink!25}
         &$D\oplus \widetilde{D}\oplus E$&$ 4$&\colorbox{pink!25}{\makecell*{$D\widetilde{D}D\widetilde{D}$,$D\widetilde{D}\phi^2$}}\\
          \rowcolor{pink!25}
         &$L\oplus E\oplus \widetilde{D}$&$ 4$&\colorbox{pink!25}{\makecell*{$LLE\phi$, $LLE\phi^2$}}\\
          \rowcolor{pink!25}
          &$L\oplus \widetilde{L}\oplus E$&$ 4$&\colorbox{pink!25}{\makecell*{$LLE\phi $, $L\widetilde{L}L\widetilde{L}$, $L\widetilde{L}\phi^2$ }}\\
          \rowcolor{pink!25}
        \multirow{-7}*{$\five \oplus \fivebar \oplus \mathbf{10}$} &$Q\oplus\widetilde{D}\oplus L$&$ 3$&$QQ\widetilde{D}$\\
         \hline  
           \rowcolor{purple!25}
           &$Q \oplus \widetilde{D} \oplus \widetilde{U}$& $ 3$ &\colorbox{purple!25}{\makecell*{$QQ\widetilde{D}$, $\widetilde{D}\widetilde{D}\widetilde{U}$}}\\
          \rowcolor{purple!25}
         &$\widetilde{D}\oplus \widetilde{U} \oplus \widetilde{L} \oplus Q$&$ 3$&\colorbox{purple!25}{\makecell*{$QQ\widetilde{D}$,  $\widetilde{D}\widetilde{D}\widetilde{U}$}}\\
          \rowcolor{purple!25}
         &$Q\oplus \widetilde{D} \oplus \widetilde{E}$& $ 3$& $QQ\widetilde{D}$\\
          \rowcolor{purple!25}
         \multirow{-5}*{$\five\oplus \mathbf{10}\oplus\mathbf{\overline{10}}$}&$\widetilde{D}\oplus \widetilde{U} \oplus E$& $3$& $\widetilde{D}\widetilde{D}\widetilde{U}$\\
         \hline
         \rowcolor{blue!25}        
        &$Q\oplus \widetilde{D}\oplus L\oplus\widetilde{U}$& $ 3$& \colorbox{blue!25}{\makecell*{$QQ\widetilde{D}$, $\widetilde{D}\widetilde{D}\widetilde{U}$}}\\
           \rowcolor{blue!25}    
           \multirow{-2}*{$\fivebar \oplus \mathbf{\overline{10}} \oplus \five \oplus \mathbf{10}$}&$Q\oplus \widetilde{D} \oplus L \oplus \widetilde{E}$& $   3$& $QQ\widetilde{D}$\\
         \hline
         \end{tabular}
\end{table}

In Tab. \ref{tab:all-models} and Tab. \ref{tab:all-models-cont}, we show all 
the models that we found to be viable, with the exclusion of those models in which $N$ is light. Remember that exchange each fermion with its tilded version 
$$\Psi \leftrightarrow \widetilde{\Psi}$$ 
results in another viable model with the same features. 
The DMC is taken to be the least charged DB with smallest spin. 

Generally, one can extend these models with singlet fermions $N$ if one makes sure that the DB number remains unbroken. Essentially this boils down to excluding $\NDC=3$, which is spoiled by the dimension five operators of Eq. \eqref{eq:Nbreak3colors}. Unless the Dq $L$ is light, the $N$ species number is only broken at dimension five leading to metastable mesons. 


%% file: Bibliography.bib
@article{unificaxion,
    author = "Giudice, G. F. and Rattazzi, R. and Strumia, A.",
    title = "{Unificaxion}",
    eprint = "1204.5465",
    archivePrefix = "arXiv",
    primaryClass = "hep-ph",
    reportNumber = "CERN-PH-TH-2012-098",
    doi = "10.1016/j.physletb.2012.07.028",
    journal = "Phys. Lett. B",
    volume = "715",
    pages = "142--148",
    year = "2012"
}

@article{SU5GUT,
  title = {Unity of All Elementary-Particle Forces},
  author = {Georgi, Howard and Glashow, S. L.},
  journal = {Phys. Rev. Lett.},
  volume = {32},
  issue = {8},
  pages = {438--441},
  numpages = {0},
  year = {1974},
  %month = {February},
  publisher = {American Physical Society},
  doi = {10.1103/PhysRevLett.32.438},
  url = {https://link.aps.org/doi/10.1103/PhysRevLett.32.438}
}

@article{verma,
    author = "Verma, Sonali and Bottaro, Salvatore and Contino, Roberto",
    title = "{Accidental composite dark matter in $SU$(5)-GUT theories}",
    doi = "10.22323/1.414.0306",
    journal = "PoS",
    volume = "ICHEP2022",
    pages = "306",
    month = "11",
    year = "2022"
}

@article{WCD, 
      title={Dark Matter as a weakly coupled Dark Baryon}, 
      author={A. Mitridate M. Redi J. Smironv A. Strumia},
      year={2017},
      eprint={1707.05380},
      archivePrefix={arXiv},
      primaryClass={hep-ph}
}

@misc{CDM,
      title={Accidental Composite Dark Matter}, 
      author={O. Antipin M. Redi A. Strumia E. Vigiani},
      year={2015},
      eprint={1503.08749},
      archivePrefix={arXiv},
      primaryClass={hep-ph}
}

@misc{VLC,
      title={Vector Like Confinement at LHC}, 
      author={Can Kilic and Takemichi Okui and Raman Sundrum},
      year={2017},
      eprint={0906.0577v4},
      archivePrefix={arXiv},
      primaryClass={hep-ph}
}

@article{strumia-scalardm,
    author = "Buttazzo, Dario and Di Luzio, Luca and Landini, Giacomo and Strumia, Alessandro and Teresi, Daniele",
    title = "{Dark Matter from self-dual gauge/Higgs dynamics}",
    eprint = "1907.11228",
    archivePrefix = "arXiv",
    primaryClass = "hep-ph",
    doi = "10.1007/JHEP10(2019)067",
    journal = "JHEP",
    volume = "10",
    pages = "067",
    year = "2019"
}

@article{Morningstar_1999,
	doi = {10.1103/physrevd.60.034509},
	url = {https://doi.org/10.1103%2Fphysrevd.60.034509},
	year = 1999,
	month = {7},
	publisher = {American Physical Society ({APS})},
	volume = {60},
	number = {3},
	author = {Colin J. Morningstar and Mike Peardon},
	title = {Glueball spectrum from an anisotropic lattice study},
	journal = {Physical Review D}
}

@article{ParticleDataGroup:2020ssz,
    author = "Zyla, P. A. and others",
    collaboration = "Particle Data Group",
    title = "{Review of Particle Physics}",
    doi = "10.1093/ptep/ptaa104",
    journal = "PTEP",
    volume = "2020",
    number = "8",
    pages = "083C01",
    year = "2020"
}

@article{MDM,
    author = "Cirelli, Marco and Fornengo, Nicolao and Strumia, Alessandro",
    title = "{Minimal dark matter}",
    eprint = "hep-ph/0512090",
    archivePrefix = "arXiv",
    reportNumber = "DFTT40-2005, IFUP-TH-2005-34",
    doi = "10.1016/j.nuclphysb.2006.07.012",
    journal = "Nucl. Phys. B",
    volume = "753",
    pages = "178--194",
    year = "2006"
}

@article{darkmesonsLHC,
    author = "Kribs, Graham D. and Martin, Adam and Ostdiek, Bryan and Tong, Tom",
    title = "{Dark Mesons at the LHC}",
    eprint = "1809.10184",
    archivePrefix = "arXiv",
    primaryClass = "hep-ph",
    doi = "10.1007/JHEP07(2019)133",
    journal = "JHEP",
    volume = "07",
    pages = "133",
    year = "2019"
}

@article{ATLAS:darkmesons,
    collaboration = "ATLAS",
    title = "{Search for dark mesons decaying to top and bottom quarks with the ATLAS detector in 140 fb$^{-1}$ of proton-proton collisions at $\sqrt{s}=13~$TeV}",
    reportNumber = "ATLAS-CONF-2023-021",
    year = "2023"
}

@article{Dondi:2020,
    author = "Dondi, Nicola Andrea and Sannino, Francesco and Smirnov, Juri",
    title = "{Thermal history of composite dark matter}",
    eprint = "1905.08810",
    archivePrefix = "arXiv",
    primaryClass = "hep-ph",
    doi = "10.1103/PhysRevD.101.103010",
    journal = "Phys. Rev. D",
    volume = "101",
    number = "10",
    pages = "103010",
    year = "2020"
}

@article{gluequark,
    author = "Contino, Roberto and Mitridate, Andrea and Podo, Alessandro and Redi, Michele",
    title = "{Gluequark Dark Matter}",
    eprint = "1811.06975",
    archivePrefix = "arXiv",
    primaryClass = "hep-ph",
    doi = "10.1007/JHEP02(2019)187",
    journal = "JHEP",
    volume = "02",
    pages = "187",
    year = "2019"
}

@article{phasesofcannibal,
    author = "Farina, Marco and Pappadopulo, Duccio and Ruderman, Joshua T. and Trevisan, Gabriele",
    title = "{Phases of Cannibal Dark Matter}",
    eprint = "1607.03108",
    archivePrefix = "arXiv",
    primaryClass = "hep-ph",
    doi = "10.1007/JHEP12(2016)039",
    journal = "JHEP",
    volume = "12",
    pages = "039",
    year = "2016"
}

@article{cannibal,
	doi = {10.1103/physrevd.94.035005},
  
	url = {https://doi.org/10.1103\%2Fphysrevd.94.035005},
  
	year = 2016,
	month = {8},
  
	publisher = {American Physical Society ({APS})},
  
	volume = {94},
  
	number = {3},
  
	author = {Duccio Pappadopulo and Joshua T. Ruderman and Gabriele Trevisan},
  
	title = {Dark matter freeze-out in a nonrelativistic sector},
  
	journal = {Physical Review D}
}

@article{LQS1S3,
    author = "I.Dorsner, S. Fafjer, O.Sumensari",
    title = "{Muon $g-2$ and scalar leptoquark mixing}",
    eprint = "1910.03877v2",
    archivePrefix = "arXiv",
    primaryClass = "hep-ph",
    month = "5",
    year = "2020"
}

@article{LZ,
    author = "Aalbers, J. and others",
    collaboration = "LZ",
    title = "{First Dark Matter Search Results from the LUX-ZEPLIN (LZ) Experiment}",
    eprint = "2207.03764",
    archivePrefix = "arXiv",
    primaryClass = "hep-ex",
    month = "7",
    year = "2022"
}

@article{lastWIMP,
    author = "Bottaro, Salvatore and Buttazzo, Dario and Costa, Marco and Franceschini, Roberto and Panci, Paolo and Redigolo, Diego and Vittorio, Ludovico",
    title = "{The last complex WIMPs standing}",
    eprint = "2205.04486",
    archivePrefix = "arXiv",
    primaryClass = "hep-ph",
    reportNumber = "CERN-TH-2022-080",
    doi = "10.1140/epjc/s10052-022-10918-5",
    journal = "Eur. Phys. J. C",
    volume = "82",
    number = "11",
    pages = "992",
    year = "2022"
}

@article{VLC2,
    author = "Kilic, Can and Okui, Takemichi",
    title = "{The LHC Phenomenology of Vectorlike Confinement}",
    eprint = "1001.4526",
    archivePrefix = "arXiv",
    primaryClass = "hep-ph",
    reportNumber = "RUNHETC-2010-01",
    doi = "10.1007/JHEP04(2010)128",
    journal = "JHEP",
    volume = "04",
    pages = "128",
    year = "2010"
}

@article{Planck:2018,
	doi = {10.1051/0004-6361/201833910},
  
	url = {https://doi.org/10.1051%2F0004-6361%2F201833910},
  
	year = 2020,
	month = {9},
  
	publisher = {{EDP} Sciences},
  
	volume = {641},
  
	pages = {A6},
  
	author = {N. Aghanim and Y. Akrami and M. Ashdown and J. Aumont and C. Baccigalupi and M. Ballardini and A. J. Banday and R. B. Barreiro and N. Bartolo and S. Basak and R. Battye and K. Benabed and J.-P. Bernard and M. Bersanelli and P. Bielewicz and J. J. Bock and J. R. Bond and J. Borrill and F. R. Bouchet and F. Boulanger and M. Bucher and C. Burigana and R. C. Butler and E. Calabrese and J.-F. Cardoso and J. Carron and A. Challinor and H. C. Chiang and J. Chluba and L. P. L. Colombo and C. Combet and D. Contreras and B. P. Crill and F. Cuttaia and P. de Bernardis and G. de Zotti and J. Delabrouille and J.-M. Delouis and E. Di Valentino and J. M. Diego and O. Dor{\'{e}
} and M. Douspis and A. Ducout and X. Dupac and S. Dusini and G. Efstathiou and F. Elsner and T. A. En{\ss}lin and H. K. Eriksen and Y. Fantaye and M. Farhang and J. Fergusson and R. Fernandez-Cobos and F. Finelli and F. Forastieri and M. Frailis and A. A. Fraisse and E. Franceschi and A. Frolov and S. Galeotta and S. Galli and K. Ganga and R. T. G{\'{e}}nova-Santos and M. Gerbino and T. Ghosh and J. Gonz{\'{a}}lez-Nuevo and K. M. G{\'{o}}rski and S. Gratton and A. Gruppuso and J. E. Gudmundsson and J. Hamann and W. Handley and F. K. Hansen and D. Herranz and S. R. Hildebrandt and E. Hivon and Z. Huang and A. H. Jaffe and W. C. Jones and A. Karakci and E. Keihänen and R. Keskitalo and K. Kiiveri and J. Kim and T. S. Kisner and L. Knox and N. Krachmalnicoff and M. Kunz and H. Kurki-Suonio and G. Lagache and J.-M. Lamarre and A. Lasenby and M. Lattanzi and C. R. Lawrence and M. Le Jeune and P. Lemos and J. Lesgourgues and F. Levrier and A. Lewis and M. Liguori and P. B. Lilje and M. Lilley and V. Lindholm and M. L{\'{o}}pez-Caniego and P. M. Lubin and Y.-Z. Ma and J. F. Mac{\'{\i}}as-P{\'{e}}rez and G. Maggio and D. Maino and N. Mandolesi and A. Mangilli and A. Marcos-Caballero and M. Maris and P. G. Martin and M. Martinelli and E. Mart{\'{\i}}nez-Gonz{\'{a}}lez and S. Matarrese and N. Mauri and J. D. McEwen and P. R. Meinhold and A. Melchiorri and A. Mennella and M. Migliaccio and M. Millea and S. Mitra and M.-A. Miville-Desch{\^{e}}nes and D. Molinari and L. Montier and G. Morgante and A. Moss and P. Natoli and H. U. N{\o}rgaard-Nielsen and L. Pagano and D. Paoletti and B. Partridge and G. Patanchon and H. V. Peiris and F. Perrotta and V. Pettorino and F. Piacentini and L. Polastri and G. Polenta and J.-L. Puget and J. P. Rachen and M. Reinecke and M. Remazeilles and A. Renzi and G. Rocha and C. Rosset and G. Roudier and J. A. Rubi{\~{n}}o-Mart{\'{\i}}n and B. Ruiz-Granados and L. Salvati and M. Sandri and M. Savelainen and D. Scott and E. P. S. Shellard and C. Sirignano and G. Sirri and L. D. Spencer and R. Sunyaev and A.-S. Suur-Uski and J. A. Tauber and D. Tavagnacco and M. Tenti and L. Toffolatti and M. Tomasi and T. Trombetti and L. Valenziano and J. Valiviita and B. Van Tent and L. Vibert and P. Vielva and F. Villa and N. Vittorio and B. D. Wandelt and I. K. Wehus and M. White and S. D. M. White and A. Zacchei and A. Zonca},
  
	title = {Planck 2018 results},
  
	journal = {Astronomy {\&} Astrophysics}
}

@article{SUSYDM,
	doi = {10.1016/0370-1573(95)00058-5},
  
	url = {https://doi.org/10.1016%2F0370-1573%2895%2900058-5},
  
	year = 1996,
	month = {3},
  
	publisher = {Elsevier {BV}
},
  
	volume = {267},
  
	number = {5-6},
  
	pages = {195--373},
  
	author = {Gerard Jungman and Marc Kamionkowski and Kim Griest},
  
	title = {Supersymmetric dark matter},
  
	journal = {Physics Reports}
}

@article{PBH,
    author = "Sasaki, Misao and Suyama, Teruaki and Tanaka, Takahiro and Yokoyama, Shuichiro",
    title = "{Primordial black holes\textemdash{}perspectives in gravitational wave astronomy}",
    eprint = "1801.05235",
    archivePrefix = "arXiv",
    primaryClass = "astro-ph.CO",
    doi = "10.1088/1361-6382/aaa7b4",
    journal = "Class. Quant. Grav.",
    volume = "35",
    number = "6",
    pages = "063001",
    year = "2018"
}

@article{MOND,
    author = "Milgrom, Mordehai",
    title = "{MOND: A pedagogical review}",
    eprint = "astro-ph/0112069",
    archivePrefix = "arXiv",
    journal = "Acta Phys. Polon. B",
    volume = "32",
    pages = "3613",
    year = "2001"
}

@article{BBN-glueballs,
	doi = {10.1103/physrevd.74.103509},
  
	url = {https://doi.org/10.1103%2Fphysrevd.74.103509},
  
	year = 2006,
	month = {11},
  
	publisher = {American Physical Society ({APS})},
  
	volume = {74},
  
	number = {10},
  
	author = {Karsten Jedamzik},
  
	title = {Big bang nucleosynthesis constraints on hadronically and electromagnetically decaying relic neutral particles},
  
	journal = {Physical Review D}
}

@article{gammaray-glueballs,
    author = "Kribs, Graham D. and Rothstein, I. Z.",
    title = "{Bounds on longlived relics from diffuse gamma-ray observations}",
    eprint = "hep-ph/9610468",
    archivePrefix = "arXiv",
    reportNumber = "UCSD-TH-96-27",
    doi = "10.1103/PhysRevD.56.1822",
    journal = "Phys. Rev. D",
    volume = "55",
    pages = "4435--4449",
    year = "1997",
    note = "[Erratum: Phys.Rev.D 56, 1822 (1997)]"
}

@article{coloredDM,
	doi = {10.1103/physrevd.97.115024},
  
	url = {https://doi.org/10.1103%2Fphysrevd.97.115024},
  
	year = 2018,
	month = {6},
  
	publisher = {American Physical Society ({APS})},
  
	volume = {97},
  
	number = {11},
  
	author = {Valerio De Luca and Andrea Mitridate and Michele Redi and Juri Smirnov and Alessandro Strumia},
  
	title = {Colored dark matter},
  
	journal = {Physical Review D}
}

@article{costa:asymmetric-dm,
    author = "Bottaro, Salvatore and Costa, Marco and Popov, Oleg",
    title = "{Asymmetric accidental composite dark matter}",
    eprint = "2104.14244",
    archivePrefix = "arXiv",
    primaryClass = "hep-ph",
    doi = "10.1007/JHEP11(2021)055",
    journal = "JHEP",
    volume = "11",
    pages = "055",
    year = "2021"
}

@article{bearable-compositeness ,
    author = "Frigerio, Michele and Nardecchia, Marco and Serra, Javi and Vecchi, Luca",
    title = "{The Bearable Compositeness of Leptons}",
    eprint = "1807.04279",
    archivePrefix = "arXiv",
    primaryClass = "hep-ph",
    reportNumber = "CERN-TH-2018-160",
    doi = "10.1007/JHEP10(2018)017",
    journal = "JHEP",
    volume = "10",
    pages = "017",
    year = "2018"
}

@article{Grzadkowski_2010,
	doi = {10.1007/jhep10(2010)085},
  
	url = {https://doi.org/10.1007%2Fjhep10%282010%29085},
  
	year = 2010,
	month = {10},
  
	publisher = {Springer Science and Business Media {LLC}
},
  
	volume = {2010},
  
	number = {10},
  
	author = {B. Grzadkowski and M. Iskrzy{\'{n}}ski and M. Misiak and J. Rosiek},
  
	title = {Dimension-six terms in the Standard Model Lagrangian},
  
	journal = {Journal of High Energy Physics}
}

@article{Fradkin-Shenker,
    author = "Fradkin, Eduardo H. and Shenker, Stephen H.",
    title = "{Phase Diagrams of Lattice G8e Theories with Higgs Fields}",
    reportNumber = "SLAC-PUB-2238",
    doi = "10.1103/PhysRevD.19.3682",
    journal = "Phys. Rev. D",
    volume = "19",
    pages = "3682--3697",
    year = "1979"
}

@inproceedings{tHooft-duality,
    author = "'t Hooft, Gerard",
    title = "{Topological aspects of quantum chromodynamics}",
    booktitle = "{International School of Nuclear Physics: 20th Course: Heavy Ion Collisions from Nuclear to Quark Matter (Erice 98)}",
    eprint = "hep-th/9812204",
    archivePrefix = "arXiv",
    reportNumber = "SPIN-1998-19",
    pages = "216--236",
    month = "9",
    year = "1998"
}

@misc{Mathematica,
  author = {Wolfram Research{,} Inc.},
  title = {Mathematica, {V}ersion 13.3},
  url = {https://www.wolfram.com/mathematica},
  note = {Champaign, IL, 2023}
}

@article{LieART,
	doi = {10.1016/j.cpc.2020.107490},
  
	url = {https://doi.org/10.1016%2Fj.cpc.2020.107490},
  
	year = 2020,
	month = {12},
  
	publisher = {Elsevier {BV}
},
  
	volume = {257},
  
	pages = {107490},
  
	author = {Robert Feger and Thomas W. Kephart and Robert J. Saskowski},
  
	title = {{LieART} 2.0 {\textendash} A Mathematica application for Lie Algebras and Representation Theory},
  
	journal = {Computer Physics Communications}
}

@article{Contino:2020,
    author = "Contino, Roberto and Podo, Alessandro and Revello, Filippo",
    title = "{Composite Dark Matter from Strongly-Interacting Chiral Dynamics}",
    eprint = "2008.10607",
    archivePrefix = "arXiv",
    primaryClass = "hep-ph",
    doi = "10.1007/JHEP02(2021)091",
    journal = "JHEP",
    volume = "02",
    pages = "091",
    year = "2021"
}

@article{dmnaturalness,
	doi = {10.1007/jhep12(2019)037},
  
	url = {https://doi.org/10.1007%2Fjhep12%282019%29037},
  
	year = 2019,
	month = {12},
  
	publisher = {Springer Science and Business Media {LLC}
},
  
	volume = {2019},
  
	number = {12},
  
	author = {Mark P. Hertzberg and McCullen Sandora},
  
	title = {Dark matter and naturalness},
  
	journal = {Journal of High Energy Physics}
}

@article{Harigaya,
    author = "Harigaya, Keisuke and Nomura, Yasunori",
    title = "{Light Chiral Dark Sector}",
    eprint = "1603.03430",
    archivePrefix = "arXiv",
    primaryClass = "hep-ph",
    doi = "10.1103/PhysRevD.94.035013",
    journal = "Phys. Rev. D",
    volume = "94",
    number = "3",
    pages = "035013",
    year = "2016"
}

@article{LPref,
    author = "I. Doršner and S. Fajfer and A. Greljo and J. F. Kamenik and N. Košnik",
    title = "{Physics of leptoquarks in precision experiments and at particle colliders}",
    eprint = "1603.04993",
    archivePrefix = "arXiv",
    primaryClass = "hep-ph",
    month = "8",
    year = "2016"
 }

@article{id-cohen,
	doi = {10.1103/physrevlett.119.021102},
  
	url = {https://doi.org/10.1103%2Fphysrevlett.119.021102},
  
	year = 2017,
	month = {7},
  
	publisher = {American Physical Society ({APS})},
  
	volume = {119},
  
	number = {2},
  
	author = {Timothy Cohen and Kohta Murase and Nicholas L. Rodd and Benjamin R. Safdi and Yotam Soreq},
  
	title = {Gamma-ray Constraints on Decaying Dark Matter and Implications for {IceCube}
},
  
	journal = {Physical Review Letters}
}

@misc{axiondm,
      title={Axion Dark Matter}, 
      author={C. B. Adams and N. Aggarwal and A. Agrawal and R. Balafendiev and C. Bartram and M. Baryakhtar and H. Bekker and P. Belov and K. K. Berggren and A. Berlin and C. Boutan and D. Bowring and D. Budker and A. Caldwell and P. Carenza and G. Carosi and R. Cervantes and S. S. Chakrabarty and S. Chaudhuri and T. Y. Chen and S. Cheong and A. Chou and R. T. Co and J. Conrad and D. Croon and R. T. D'Agnolo and M. Demarteau and N. DePorzio and M. Descalle and K. Desch and L. Di Luzio and A. Diaz-Morcillo and K. Dona and I. S. Drachnev and A. Droster and N. Du and K. Dunne and B. Döbrich and S. A. R. Ellis and R. Essig and J. Fan and J. W. Foster and J. T. Fry and A. Gallo Rosso and J. M. García Barceló and I. G. Irastorza and S. Gardner and A. A. Geraci and S. Ghosh and B. Giaccone and M. Giannotti and B. Gimeno and D. Grin and H. Grote and M. Guzzetti and M. H. Awida and R. Henning and S. Hoof and G. Hoshino and V. Irsic and K. D. Irwin and H. Jackson and D. F. Jackson Kimball and J. Jaeckel and K. Jakovcic and M. J. Jewell and M. Kagan and Y. Kahn and R. Khatiwada and S. Knirck and T. Kovachy and P. Krueger and S. E. Kuenstner and N. A. Kurinsky and R. K. Leane and A. F. Leder and C. Lee and K. W. Lehnert and E. W. Lentz and S. M. Lewis and J. Liu and M. Lynn and B. Majorovits and D. J. E. Marsh and R. H. Maruyama and B. T. McAllister and A. J. Millar and D. W. Miller and J. Mitchell and S. Morampudi and G. Mueller and S. Nagaitsev and E. Nardi and O. Noroozian and C. A. J. O'Hare and N. S. Oblath and J. L. Ouellet and K. M. W. Pappas and H. V. Peiris and K. Perez and A. Phipps and M. J. Pivovaroff and P. Quílez and N. M. Rapidis and V. H. Robles and K. K. Rogers and J. Rudolph and J. Ruz and G. Rybka and M. Safdari and B. R. Safdi and M. S. Safronova and C. P. Salemi and P. Schuster and A. Schwartzman and J. Shu and M. Simanovskaia and J. Singh and S. Singh and K. Sinha and J. T. Sinnis and M. Siodlaczek and M. S. Smith and W. M. Snow and A. V. Sokolov and A. Sonnenschein and D. H. Speller and Y. V. Stadnik and C. Sun and A. O. Sushkov and T. M. P. Tait and V. Takhistov and D. B. Tanner and F. Tavecchio and D. J. Temples and J. H. Thomas and M. E. Tobar and N. Toro and Y. -D. Tsai and E. C. van Assendelft and K. van Bibber and M. Vandegar and L. Visinelli and E. Vitagliano and J. K. Vogel and Z. Wang and A. Wickenbrock and L. Winslow and S. Withington and M. Wooten and J. Yang and B. A. Young and F. Yu and K. Zhou and T. Zhou},
      year={2023},
      eprint={2203.14923},
      archivePrefix={arXiv},
      primaryClass={hep-ex}
}

@article{dim-raby-suss,
    author = "Dimopoulos, S. and Raby, S. and Susskind, Leonard",
    title = "{Light Composite Fermions}",
    reportNumber = "ITP-662-STANFORD",
    doi = "10.1016/0550-3213(80)90215-1",
    journal = "Nucl. Phys. B",
    volume = "173",
    pages = "208--228",
    year = "1980"
}

@article{stealthdm,
   title={Stealth dark matter: Dark scalar baryons through the Higgs portal},
   volume={92},
   ISSN={1550-2368},
   url={http://dx.doi.org/10.1103/PhysRevD.92.075030},
   DOI={10.1103/physrevd.92.075030},
   number={7},
   journal={Physical Review D},
   publisher={American Physical Society (APS)},
   author={Appelquist, T. and Brower, R. C. and Buchoff, M. I. and Fleming, G. T. and Jin, X.-Y. and Kiskis, J. and Kribs, G. D. and Neil, E. T. and Osborn, J. C. and Rebbi, C. and Rinaldi, E. and Schaich, D. and Schroeder, C. and Syritsyn, S. and Vranas, P. and Weinberg, E. and Witzel, O.},
   year={2015},
   month={10} 
}

@article{mucoll,
    author = "Delahaye, Jean Pierre and Diemoz, Marcella and Long, Ken and Mansouli\'e, Bruno and Pastrone, Nadia and Rivkin, Lenny and Schulte, Daniel and Skrinsky, Alexander and Wulzer, Andrea",
    title = "{Muon Colliders}",
    eprint = "1901.06150",
    archivePrefix = "arXiv",
    primaryClass = "physics.acc-ph",
    month = "1",
    year = "2019"
}

@article{Aartsen_2018,
   title={Search for neutrinos from decaying dark matter with IceCube: IceCube Collaboration},
   volume={78},
   ISSN={1434-6052},
   url={http://dx.doi.org/10.1140/epjc/s10052-018-6273-3},
   DOI={10.1140/epjc/s10052-018-6273-3},
   number={10},
   journal={The European Physical Journal C},
   publisher={Springer Science and Business Media LLC},
   author={Aartsen, M. G. and Ackermann, M. and Adams, J. and Aguilar, J. A. and Ahlers, M. and Ahrens, M. and Samarai, I. Al and Altmann, D. and Andeen, K. and Anderson, T. and Ansseau, I. and Anton, G. and Argüelles, C. and Auffenberg, J. and Axani, S. and Backes, P. and Bagherpour, H. and Bai, X. and Barron, J. P. and Barwick, S. W. and Baum, V. and Bay, R. and Beatty, J. J. and Becker Tjus, J. and Becker, K.-H. and BenZvi, S. and Berley, D. and Bernardini, E. and Besson, D. Z. and Binder, G. and Bindig, D. and Blaufuss, E. and Blot, S. and Bohm, C. and Börner, M. and Bos, F. and Böser, S. and Botner, O. and Bourbeau, E. and Bourbeau, J. and Bradascio, F. and Braun, J. and Brenzke, M. and Bretz, H.-P. and Bron, S. and Brostean-Kaiser, J. and Burgman, A. and Busse, R. S. and Carver, T. and Cheung, E. and Chirkin, D. and Christov, A. and Clark, K. and Classen, L. and Collin, G. H. and Conrad, J. M. and Coppin, P. and Correa, P. and Cowen, D. F. and Cross, R. and Dave, P. and Day, M. and de André, J. P. A. M. and De Clercq, C. and DeLaunay, J. J. and Dembinski, H. and De Ridder, S. and Desiati, P. and de Vries, K. D. and de Wasseige, G. and de With, M. and DeYoung, T. and Díaz-Vélez, J. C. and di Lorenzo, V. and Dujmovic, H. and Dumm, J. P. and Dunkman, M. and Dvorak, E. and Eberhardt, B. and Ehrhardt, T. and Eichmann, B. and Eller, P. and Evenson, P. A. and Fahey, S. and Fazely, A. R. and Felde, J. and Filimonov, K. and Finley, C. and Flis, S. and Franckowiak, A. and Friedman, E. and Fritz, A. and Gaisser, T. K. and Gallagher, J. and Ganster, E. and Gerhardt, L. and Ghorbani, K. and Giang, W. and Glauch, T. and Glüsenkamp, T. and Goldschmidt, A. and Gonzalez, J. G. and Grant, D. and Griffith, Z. and Haack, C. and Hallgren, A. and Halve, L. and Halzen, F. and Hanson, K. and Hebecker, D. and Heereman, D. and Helbing, K. and Hellauer, R. and Hickford, S. and Hignight, J. and Hill, G. C. and Hoffman, K. D. and Hoffmann, R. and Hoinka, T. and Hokanson-Fasig, B. and Hoshina, K. and Huang, F. and Huber, M. and Hultqvist, K. and Hünnefeld, M. and Hussain, R. and In, S. and Iovine, N. and Ishihara, A. and Jacobi, E. and Japaridze, G. S. and Jeong, M. and Jero, K. and Jones, B. J. P. and Kalaczynski, P. and Kang, W. and Kappes, A. and Kappesser, D. and Karg, T. and Karle, A. and Katz, U. and Kauer, M. and Keivani, A. and Kelley, J. L. and Kheirandish, A. and Kim, J. and Kim, M. and Kintscher, T. and Kiryluk, J. and Kittler, T. and Klein, S. R. and Koirala, R. and Kolanoski, H. and Köpke, L. and Kopper, C. and Kopper, S. and Koschinsky, J. P. and Koskinen, D. J. and Kowalski, M. and Krings, K. and Kroll, M. and Krückl, G. and Kunwar, S. and Kurahashi, N. and Kuwabara, T. and Kyriacou, A. and Labare, M. and Lanfranchi, J. L. and Larson, M. J. and Lauber, F. and Leonard, K. and Lesiak-Bzdak, M. and Leuermann, M. and Liu, Q. R. and Lohfink, E. and Mariscal, C. J. Lozano and Lu, L. and Lünemann, J. and Luszczak, W. and Madsen, J. and Maggi, G. and Mahn, K. B. M. and Mancina, S. and Maruyama, R. and Mase, K. and Maunu, R. and Meagher, K. and Medici, M. and Meier, M. and Menne, T. and Merino, G. and Meures, T. and Miarecki, S. and Micallef, J. and Momenté, G. and Montaruli, T. and Moore, R. W. and Moulai, M. and Nahnhauer, R. and Nakarmi, P. and Naumann, U. and Neer, G. and Niederhausen, H. and Nowicki, S. C. and Nygren, D. R. and Obertacke Pollmann, A. and Olivas, A. and O’Murchadha, A. and O’Sullivan, E. and Palczewski, T. and Pandya, H. and Pankova, D. V. and Peiffer, P. and Pepper, J. A. and Pérez de los Heros, C. and Pieloth, D. and Pinat, E. and Plum, M. and Price, P. B. and Przybylski, G. T. and Raab, C. and Rädel, L. and Rameez, M. and Rauch, L. and Rawlins, K. and Rea, I. C. and Reimann, R. and Relethford, B. and Relich, M. and Resconi, E. and Rhode, W. and Richman, M. and Robertson, S. and Rongen, M. and Rott, C. and Ruhe, T. and Ryckbosch, D. and Rysewyk, D. and Safa, I. and Sanchez Herrera, S. E. and Sandrock, A. and Sandroos, J. and Santander, M. and Sarkar, S. and Sarkar, S. and Satalecka, K. and Schaufel, M. and Schlunder, P. and Schmidt, T. and Schneider, A. and Schoenen, S. and Schöneberg, S. and Schumacher, L. and Sclafani, S. and Seckel, D. and Seunarine, S. and Soedingrekso, J. and Soldin, D. and Song, M. and Spiczak, G. M. and Spiering, C. and Stachurska, J. and Stamatikos, M. and Stanev, T. and Stasik, A. and Stein, R. and Stettner, J. and Steuer, A. and Stezelberger, T. and Stokstad, R. G. and Stößl, A. and Strotjohann, N. L. and Stuttard, T. and Sullivan, G. W. and Sutherland, M. and Taboada, I. and Tatar, J. and Tenholt, F. and Ter-Antonyan, S. and Terliuk, A. and Tilav, S. and Toale, P. A. and Tobin, M. N. and Tönnis, C. and Toscano, S. and Tosi, D. and Tselengidou, M. and Tung, C. F. and Turcati, A. and Turley, C. F. and Ty, B. and Unger, E. and Usner, M. and Vandenbroucke, J. and Van Driessche, W. and van Eijk, D. and van Eijndhoven, N. and Vanheule, S. and van Santen, J. and Vraeghe, M. and Walck, C. and Wallace, A. and Wallraff, M. and Wandler, F. D. and Wandkowsky, N. and Waza, A. and Weaver, C. and Weiss, M. J. and Wendt, C. and Werthebach, J. and Westerhoff, S. and Whelan, B. J. and Wiebe, K. and Wiebusch, C. H. and Wille, L. and Williams, D. R. and Wills, L. and Wolf, M. and Wood, J. and Wood, T. R. and Woolsey, E. and Woschnagg, K. and Wrede, G. and Xu, D. L. and Xu, X. W. and Xu, Y. and Yanez, J. P. and Yodh, G. and Yoshida, S. and Yuan, T.},
   year={2018},
   month=10 }

@misc{MEGII,
      title={A search for $\mu^+\to e^+\gamma$ with the first dataset of the MEG II experiment}, 
      author={MEG II collaboration and K. Afanaciev and A. M. Baldini and S. Ban and V. Baranov and H. Benmansour and M. Biasotti and G. Boca and P. W. Cattaneo and G. Cavoto and F. Cei and M. Chiappini and G. Chiarello and A. Corvaglia and F. Cuna and G. Dal Maso and A. De Bari and M. De Gerone and L. Ferrari Barusso and M. Francesconi and L. Galli and G. Gallucci and F. Gatti and L. Gerritzen and F. Grancagnolo and E. G. Grandoni and M. Grassi and D. N. Grigoriev and M. Hildebrandt and K. Ieki and F. Ignatov and F. Ikeda and T. Iwamoto and S. Karpov and P. -R. Kettle and N. Khomutov and S. Kobayashi and A. Kolesnikov and N. Kravchuk and V. Krylov and N. Kuchinskiy and W. Kyle and T. Libeiro and V. Malyshev and A. Matsushita and M. Meucci and S. Mihara and W. Molzon and Toshinori Mori and M. Nakao and D. Nicolò and H. Nishiguchi and A. Ochi and S. Ogawa and R. Onda and W. Ootani and A. Oya and D. Palo and M. Panareo and A. Papa and V. Pettinacci and A. Popov and F. Renga and S. Ritt and M. Rossella and A. Rozhdestvensky and P. Schwendimann and K. Shimada and G. Signorelli and M. Takahashi and G. F. Tassielli and K. Toyoda and Y. Uchiyama and M. Usami and A. Venturini and B. Vitali and C. Voena and K. Yamamoto and K. Yanai and T. Yonemoto and K. Yoshida and Yu. V. Yudin},
      year={2024},
      eprint={2310.12614},
      archivePrefix={arXiv},
      primaryClass={hep-ex}
}

@article{cms-lq,
    author = "Hayrapetyan, Aram and others",
    collaboration = "CMS",
    title = "{Search for pair production of scalar and vector leptoquarks decaying to muons and bottom quarks in proton-proton collisions at $\sqrt{s}$ = 13 TeV}",
    eprint = "2402.08668",
    archivePrefix = "arXiv",
    primaryClass = "hep-ex",
    reportNumber = "CMS-EXO-21-019, CERN-EP-2023-301",
    month = "2",
    year = "2024"
}

@article{qcd-hybrid-baryons,
    author = "Dudek, Jozef J. and Edwards, Robert G.",
    title = "{Hybrid Baryons in QCD}",
    eprint = "1201.2349",
    archivePrefix = "arXiv",
    primaryClass = "hep-ph",
    reportNumber = "JLAB-THY-12-1479",
    doi = "10.1103/PhysRevD.85.054016",
    journal = "Phys. Rev. D",
    volume = "85",
    pages = "054016",
    year = "2012"
}

@article{qcd-hybrids,
    author = "Meyer, C. A. and Swanson, E. S.",
    title = "{Hybrid Mesons}",
    eprint = "1502.07276",
    archivePrefix = "arXiv",
    primaryClass = "hep-ph",
    doi = "10.1016/j.ppnp.2015.03.001",
    journal = "Prog. Part. Nucl. Phys.",
    volume = "82",
    pages = "21--58",
    year = "2015"
}

@article{Redi:dark-nuclei,
    author = "Redi, Michele and Tesi, Andrea",
    title = "{Cosmological Production of Dark Nuclei}",
    eprint = "1812.08784",
    archivePrefix = "arXiv",
    primaryClass = "hep-ph",
    doi = "10.1007/JHEP04(2019)108",
    journal = "JHEP",
    volume = "04",
    pages = "108",
    year = "2019"
}

@article{Mahbubani:dark-nucleosynthesis,
    author = "Mahbubani, Rakhi and Redi, Michele and Tesi, Andrea",
    title = "{Dark Nucleosynthesis: Cross-sections and Astrophysical Signals}",
    eprint = "2007.07231",
    archivePrefix = "arXiv",
    primaryClass = "hep-ph",
    doi = "10.1088/1475-7516/2021/02/039",
    journal = "JCAP",
    volume = "02",
    pages = "039",
    year = "2021"
}

@article{Kribs:review-composite,
    author = "Kribs, Graham D. and Neil, Ethan T.",
    title = "{Review of strongly-coupled composite dark matter models and lattice simulations}",
    eprint = "1604.04627",
    archivePrefix = "arXiv",
    primaryClass = "hep-ph",
    doi = "10.1142/S0217751X16430041",
    journal = "Int. J. Mod. Phys. A",
    volume = "31",
    number = "22",
    pages = "1643004",
    year = "2016"
}

@article{Barr:asymmetric-dm,
    author = "Barr, Stephen M. and Chivukula, R. Sekhar and Farhi, Edward",
    title = "{Electroweak Fermion Number Violation and the Production of Stable Particles in the Early Universe}",
    reportNumber = "NSF-ITP-90-27, BA-90-7, BUHEP-90-6, MIT-CTP-1833",
    doi = "10.1016/0370-2693(90)91661-T",
    journal = "Phys. Lett. B",
    volume = "241",
    pages = "387--391",
    year = "1990"
}

@article{Song:fermilat,
    author = "Song, Deheng and Murase, Kohta and Kheirandish, Ali",
    title = "{Constraining decaying very heavy dark matter from galaxy clusters with 14 year Fermi-LAT data}",
    eprint = "2308.00589",
    archivePrefix = "arXiv",
    primaryClass = "astro-ph.HE",
    doi = "10.1088/1475-7516/2024/03/024",
    journal = "JCAP",
    volume = "03",
    pages = "024",
    year = "2024"
}

@article{HAWC:2023bti,
    author = "Albert, A. and others",
    collaboration = "HAWC",
    title = "{Search for decaying dark matter in the Virgo cluster of galaxies with HAWC}",
    eprint = "2309.03973",
    archivePrefix = "arXiv",
    primaryClass = "astro-ph.HE",
    doi = "10.1103/PhysRevD.109.043034",
    journal = "Phys. Rev. D",
    volume = "109",
    number = "4",
    pages = "043034",
    year = "2024"
}
